\documentclass[%
 reprint,
 amsmath,amssymb,
 aps,
prb,
]{revtex4-2}

\usepackage{graphicx}
\usepackage{dcolumn}
\usepackage{bm}

\usepackage{color}
\usepackage{braket} 
\usepackage{txfonts}
\usepackage{ulem}

\newcommand{\rev}[1]{\textcolor{black}{#1}}

\AtBeginDocument{\renewcommand{\natexlab}[1]{#1}}

\begin{document}

\preprint{APS/123-QED}

\title{NQR and NMR \rev{s}pectra in \rev{o}dd-\rev{p}arity \rev{m}ultipole \rev{m}aterial CeCoSi}

\author{Megumi Yatsushiro$^{1,2}$, Satoru Hayami$^{2}$}
 \affiliation{$^1$Department of Physics, Hokkaido University, Sapporo 060-0810, Japan \\
 $^2$ Department of Applied Physics, The University of Tokyo, Tokyo 113-8656, Japan}
 
\begin{abstract}
We theoretically study NQR and NMR spectra in the presence of odd-parity multipoles originating from staggered antiferromagnetic and antiferroquadrupole orderings.
For the $f$-electron metal, CeCoSi, which is a candidate hosting odd-parity multipoles,
we derive an effective hyperfine field acting on Co nucleus generated from electronic origin multipole moments at Ce ion under zero and nonzero magnetic fields.
We elucidate that emergent odd-parity multipoles give rise to sublattice-dependent spectral splittings in NQR and NMR through the effective hyperfine coupling in the absence of the global inversion symmetry.
We mainly examine behaviors of the NQR and NMR spectra in three odd-parity multipole ordered states:
a $y$-type magnetic toroidal dipole order with a staggered $x$-type antiferromagnetic structure,
an $xy$-type electric toroidal quadrupole order with a staggered $x^2-y^2$-type antiferroquadrupole structure, and a $z$-type electric dipole order with a staggered $3z^2-r^2$-type antiferroquadrupole structure.
We show that different odd-parity multipole orders lead to different field-dependent spectral splittings in NMR, while only the $xy$-type electric toroidal quadrupole order exhibits the NQR spectral splitting.
We also present possible sublattice-dependent spectral splittings for all the odd-parity multipole orders potentially activated in low-energy crystal-field levels, which will be useful to identify odd-parity order parameters in CeCoSi by NQR and NMR measurements.
\end{abstract}

\maketitle

\section{Introduction \label{sec:intro}}
Spatial inversion symmetry is one of the fundamental symmetries in solids.
In recent studies, spontaneous inversion symmetry breaking by electronic orderings have attracted much attention, as they lead to fascinating phenomena, such as magneto-electric effect~\cite{fiebig2005revival, van2008multiferroicity, tokura2014multiferroics} and nonreciprocal transport properties~\cite{tokura2018nonreciprocal}.
Once the systems undergo phase transitions causing inversion symmetry breaking, 
order parameters are represented by unconventional odd-parity multipoles, such as magnetic toroidal dipole~\cite{spaldin2008toroidal, hayami2014toroidal, hayami2015spontaneous, saito2018evidence,thole2018magnetoelectric}, magnetic quadrupole~\cite{khanh2017manipulation, yanagi2018manipulating, watanabe2017magnetic}, electric toroidal quadrupole~\cite{fu2015parity,hayami2019electric}, and electric octupole~\cite{hitomi2014electric, hitomi2016electric, hitomi2019magnetoelectric}.
In previous studies, such odd-parity multipoles have often been described by the staggered (antiferroic) alignment of even-parity multipoles 
\rev{on a crystal structure without local inversion symmetry at an atomic site; prototypes are
the zigzag chain~\cite{yanase2013magneto, hayami2015spontaneous}, honeycomb structure~\cite{hayami2014toroidal, saito2018evidence}, and diamond structure~\cite{hayami2019emergent, ishitobi2019magnetoelectric}.}
\rev{Such odd-parity multipoles
\rev{formed by} an antiferroic alignment of the even-parity multipoles like 
magnetic dipole and electric quadrupole} 
\rev{a}re denoted as cluster odd-parity multipoles.

\begin{figure}[htb!]
\centering
\includegraphics[width=88mm]{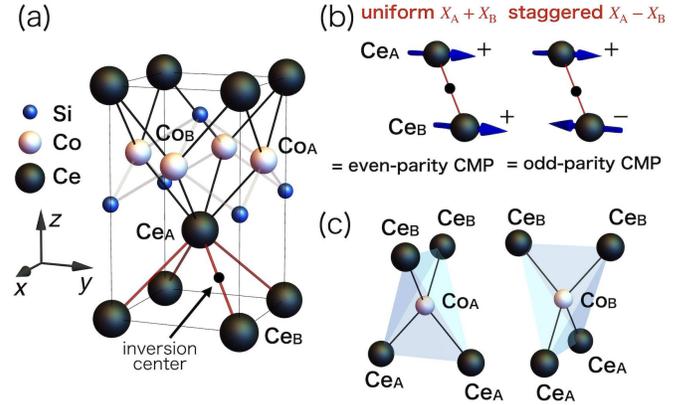}
\caption{(a) Tetragonal crystal structure of CeCoSi. 
The nearest neighbor Ce-Ce, Ce-Co, and Co-Si bonds are represented by red, black, and gray solid lines, respectively.
The rectangular represents a unit cell.
(b) Uniform and staggered alignments of \rev{local}  \rev{even-parity} multipoles in a unit cell, 
\rev{$X_{\rm A}+X_{\rm B}$} and 
\rev{$X_{\rm A}-X_{\rm B}$,} which correspond to \rev{cluster} even-parity and odd-parit\rev{y m}ultipoles (CMP) $X^{\rm (c)}$, respectively.
As an example, uniform and staggered alignments of $M_y$ are shown.
(c) Ce tetrahedrons surrounding Co$_{\rm A}$(left) and Co$_{\rm B}$(right) sites.
\label{fig:figure1}}
\end{figure}

Meanwhile, it is still an open issue how to detect cluster odd-parity multipoles. 
As emergent odd-parity multipoles are a source of physical phenomena related to inversion symmetry breaking as mentioned above, the presence/absence of odd-parity multipoles
can be distinguishable through macroscopic measurements. 
However, it is difficult to obtain microscopic information with respect to odd-parity multipoles from those measurements,
because measured physical quantities are sensitively affected by
various factors, such as domain distributions and electronic band structures. 
Thus, using probes to directly detect odd-parity multipoles are promising, such as second harmonic generation~\cite{van2007observation, zimmermann2014ferroic} and magneto-electric optics~\cite{alagna1998x, goulon1998x, goulon1999x, goulon2000x, goulon2000first, goulon2002x, kubota2004x, arima2005resonant, kimura2020imaging}. 
For example, the second harmonic generation enables us to detect the domain structure of odd-parity magnetic quadrupole and magnetic toroidal dipole and the resonant magneto-electric x-ray scattering provides the temperature dependence of the odd-parity magnetic toroidal dipole moment.

The NQR and NMR measurements are also sensitive microscopic probe to detect electronic multipoles through nuclear spins, which have been developed for exploring atomic-scale electric quadrupole and magnetic dipole/octupole in the localized $f$-electron materials, such as Ce$_{1-x}$La$_x$B$_6$~\cite{kawakami198111b,kawakami1982magnetic,kawakami1983nmr,takigawa1983nmr,sakai1997new, shiina1998interplay, sakai1999antiferro, hanzawa1999comments, hanzawa2000hyperfine,tsuji2001analysis2,tsuji2001analysis3,magishi200211b,sakai2005invariant}, NpO$_2$~\cite{sakai2005invariant,tokunaga2005nmr,tokunaga2006nmr}, and skutterudite $RT_4X_{12}$($R$: rare earth, $T$: transition metal, $X$: pnictogen)~\cite{ishida2005static,kikuchi2005quadrupole,sakai2007experimental,kikuchi2007symmetry}. 
However, the studies by the NQR and NMR measurements have been limited to ``even-parity" multipoles with respect to the spatial inversion operation, as nuclear spins and their products are characterized as even-parity tensors.

In the present study, we propose that the NQR and NMR can be another good probe to detect cluster odd-parity multipoles.
To demonstrate that, we analyze the NQR and NMR spectra under the odd-parity multipole orderings by considering the candidate, CeCoSi, which may host two types of odd-parity multipole orders depending on temperature and pressure~\cite{chevalier2006antiferromagnetic, lengyel2013temperature, tanida2018substitution, tanida2019successive, kawamura2020structural, nikitin2020gradual, tanida2020magnetic, chandra2020phenomenological}. 
The crystal structure of CeCoSi is the centrosymmetric CeFeSi-type structure ($P4/nmm$, $D_{\rm 4h}^7$, No.~$129$) in Fig.~\ref{fig:figure1}(a), where there is no inversion center on each atom; Ce and Si have $C_{\rm 4v}$ symmetry and Co has $D_{\rm 2d}$ symmetry~\cite{bodak1970crystal}. 
\rev{Such a crystal structure without local inversion symmetry at an atomic site enhances the chance to realize the cluster odd-parity multipole\rev{s} by the antiferroic alignment of even-parity multipoles as mentioned above.}
\rev{In this compound, a}s 
two different Ce sites, Ce$_{\rm A}$ and Ce$_{\rm B}$, \rev{are located at the positions without the local inversion symmetry and the inversion center is present at their bond center,} 
the staggered even-parity multipole order at \rev{those} Ce sites breaks the global inversion symmetry, which corresponds to emergence of the odd-parity multipole orders~\cite{jpsj_yatsushiro2020odd,proc_yatsushiro2020antisymmetric}.
Recently, the antiferromagnetic (AFM) ordered state at low temperature was identified as a staggered alignment of the magnetic moments along the $[100]$ direction with the ordering vector ${\bm q}={\bm 0}$ by the neutron diffraction measurement~\cite{nikitin2020gradual}, whereas the other phase mainly found under pressure [refereed as pressure induced ordered phase (PIOP)] might correspond to the antiferroquadrupole (AFQ) phase, although it is the as-yet unidentified phase~\cite{tanida2018substitution, tanida2019successive}. 
Theoretically, the authors clarified that odd-parity multipole moments are induced when the staggered AFM and AFQ phases are realized~\cite{jpsj_yatsushiro2020odd}: the identified AFM state corresponds to the \rev{cluster} magnetic toroidal dipole order and the AFQ state corresponds to any of the \rev{cluster} electric toroidal quadrupole or \rev{cluster} electric dipole order depending on types of electric quadrupoles at Ce ion. 
\rev{Thus, CeCoSi is expected to exhibit electronic-order-driven noncentrosymmetric physics, such as the Edelstein effect, magneto-electric effect, and current-induced piezoelectric effect, which originate from the cluster odd-parity multipoles~\cite{jpsj_yatsushiro2020odd}.
}

For observed or suggested odd-parity multipole orderings, we derive a general form of the hyperfine field on $^{59}$Co nucleus at zero and nonzero magnetic fields by the aid of magnetic point group analysis. 
We elucidate that the Co nuclear spins are coupled with the odd-parity order parameters from Ce ions once the spatial inversion symmetry is broken by spontaneous even-parity multipole orderings.
The hyperfine couplings arising from the odd-parity multipoles induce the sublattice-dependent NQR and NMR spectral splittings.
We show that different odd-parity multipole orders give rise to different field-dependent NMR spectral splittings, e.g., the $y$-type magnetic toroidal dipole in the $x$-type AFM structure shows the sublattice-dependent splitting except for the magnetic field normal to the $[010]$ direction. 
Furthermore, we find that only the $xy$-type electric toroidal quadrupole order arising from the $x^2-y^2$-type AFQ structure induces the sublattice-dependent NQR spectral splitting.
We also show that the hyperfine fields from the odd-parity multipoles can be evaluated from the NQR and NMR splittings.
Our result indicates that the NQR and NMR spectra in noncentrosymmetric systems will not only provide information of microscopic hyperfine fields regarding odd-parity multipoles but also be useful to identify what types of odd-parity multipoles emerge.

This paper is organized as follow.
In Sec.~\ref{sec:local}, we introduce the multipole degrees of freedom and the local Hamiltonian at Ce ions.
In Sec.~\ref{sec:hf}, the effective hyperfine field acting on the Co nucleus is derived based on the symmetry discussion.
The NQR and NMR spectra in the odd-parity multipole orders are shown in Secs.~\ref{sec:NQR} and \ref{sec:NMR}, respectively.
We summarize the NQR and NMR splittings for possible odd-parity order parameters in Sec.~\ref{sec:summary_spectra}.
Section~\ref{sec:summary} is devoted to summarize this paper.
In Appendices, we discuss a molecular mean field dependence of the odd-parity multipole moments in Appendix~\ref{ap:power}, show the spectra of the field-swept NMR in Appendix~\ref{ap:NMR_H}, present the result of the $[110]$-field NMR in Appendix~\ref{ap:NMR_110}, and summarize the result for another choice of the crystal-field level in Appendix~\ref{ap:Gamma7}.

\section{Local multipole moment at $\rev{{\rm Ce}}$ 
ion \label{sec:local}}
We introduce electronic multipole degrees of freedom at Ce ions in this section.
Starting from presenting the local multipole degrees of freedom in a $4f$ electron at Ce ion in Sec.~\ref{sec:local_multipole}, we construct the local Hamiltonian in Sec.~\ref{sec:local_Hamiltonian}.
We show the behavior of the multipole moments induced by \rev{the} AFM and AFQ states in Sec.~\ref{sec:local_AF}.
 
\subsection{Multipole degrees of freedom \label{sec:local_multipole}}

We briefly review the local and cluster multipole degrees of freedom in CeCoSi followed by
Ref.~\onlinecite{jpsj_yatsushiro2020odd}. 
\rev{We take into account multipoles activated in the $J=5/2$ multiplet from the $f^1$ configuration in a Ce$^{3+}$ ion.
The sixfold degeneracy of $J=5/2$ multiplet splits into one $\Gamma_6$ level and two $\Gamma_7$ levels under the tetragonal crystal field.}
\rev{The experiments indicate that the first and second excited states are located above 100~K and 150~K from the ground state~\cite{tanida2019successive, nikitin2020gradual}.}
\rev{In the present discussion, we consider}
\rev{t}he local multipole degrees of freedom at \rev{a} Ce io\rev{n d}escribed by the low-energy crystal-field levels \rev{up to the first-excited level.} 
\rev{W}e assume that the low-energy levels consist of the ground-state $\Gamma_7$ doublet and the first-excited $\Gamma_6$ doublet in $C_{\rm 4v}$ site symmetry~\cite{jpsj_yatsushiro2020odd}.
\rev{Within these low-energy levels, even-parity electric and magnetic multipoles with rank $l \leq 5$ are activated, as discussed below~\cite{kusunose2008description, kuramoto2009multipole, santini2009multipolar, suzuki2018first, jpsj_yatsushiro2020odd}.}
\rev{We also show the result for another low-energy levels, which consist of two $\Gamma_7$ doublets, in Appendix~\ref{ap:Gamma7}.}

For the basis function, $\phi=(\phi_{\Gamma_6 \uparrow}, \phi_{\Gamma_6 \downarrow}, \phi_{\Gamma_7 \uparrow}, \phi_{\Gamma_7 \downarrow})$, where $\uparrow, \downarrow$ represent the quasi-spin,
the local multipole degrees of freedom at Ce ion are expressed as the tensor product of two Pauli matrices, $\sigma_\mu$ within the $\Gamma_6$ or $\Gamma_7$ doublet and $\tau_\mu$ between the $\Gamma_6$-$\Gamma_7$ doublets for $\mu=0,x,y,z$ ($\sigma_0$ and $\tau_0$ are the unit matrices)\rev{~\cite{ohkawa1985orbital}}. 
The total sixteen multipoles are given as follows:
 an electric monopole (charge) $\hat{Q}_0 = \frac{1}{2}\sigma_0\tau_0$, two sets of three magnetic dipoles $(\hat{M}_{x}^\Gamma, \hat{M}_{y}^\Gamma, \hat{M}_{z}^\Gamma)=\frac{1}{4}(\sigma_x, \sigma_y, \sigma_z)(\tau_0\pm\tau_z)$, where the sign is $+(-)$
for the $\Gamma=\Gamma_6 (\Gamma_7)$ level,
and an electric quadrupole $\hat{Q}_{u(=3z^2-r^2)} = \frac{1}{2}\sigma_0\tau_z$ in a Hilbert space within the $\Gamma_6$ or $\Gamma_7$ doublet,  and 
two magnetic dipoles $(\hat{M}'_{x}, \hat{M}'_{y}) = \frac{1}{2}(\sigma_x\tau_x, -\sigma_y\tau_x)$, 
four electric quadrupoles $(\hat{Q}_{\varv(=x^2-y^2)}, \hat{Q}_{xy}, \hat{Q}_{yz}, \hat{Q}_{zx}) = \frac{1}{2} (\sigma_0\tau_x, \sigma_z \tau_y, \sigma_x\tau_y,  \sigma_y\tau_y) $,
 and two magnetic octupoles  $(\hat{M}_{xyz}, \hat{M}_{z}^\beta) = \frac{1}{2} (\sigma_0 \tau_y, \sigma_z\tau_x)$ activated in a Hilbert space between the $\Gamma_6$-$\Gamma_7$ doublets.

As there are two Ce ions in a unit cell, Ce$_{\rm A}$ and Ce$_{\rm B}$, as shown in Fig.~\ref{fig:figure1}(a), order parameters without breaking the translational symmetry are described by the uniform or staggered component of multipole moments in Ce$_{\rm A}$ and Ce$_{\rm B}$ sites:
the uniform component $\rev{\hat{X}_{\rm A}+\hat{X}_{\rm B}} $ and staggered component $\rev{\hat{X}_{\rm A}-\hat{X}_{\rm B}}$, where $\hat{X}_{i}$ stands for above multipole degrees of freedom at site $i={\rm A}, {\rm B}$.
From the symmetry viewpoint, 
\rev{uniform and staggered components}
are assigned as any of \rev{cluster}
even- and odd-parit\rev{y m}ultipoles, as shown in Fig.~\ref{fig:figure1}(b)
\rev{~\cite{jpsj_yatsushiro2020odd}.} 
\rev{We adopt the lowest-rank multipoles \rev{from the four types of multipole expressions \rev{(electric, magnetic, electric toroidal, and magnetic toroidal)}} in each uniform and staggered order\rev{~\cite{hayami2018microscopic,kusunose2020complete}}, as the multipoles with a different rank belong to the same irreducible representation in a lattice system.}
For the uniform component $\rev{\hat{X}_{\rm A}+\hat{X}_{\rm B}}$, \rev{cluster} even-parit\rev{y 
m}ultipoles are defined as
an electric monopole $\rev{\hat{Q}_0^{\rm (c)}}$ \rev{when the local multipole $X$ is $\hat{Q}_0$}, 
three magnetic dipoles 
\rev{$(\hat{M}_x^{\rm (c)}, \hat{M}_y^{\rm (c)}, \hat{M}_z^{\rm (c)})$ when $X$ is
$\sum_{\Gamma} (\hat{M}_x^{\Gamma}, \hat{M}_y^{\Gamma}, \hat{M}_z^{\Gamma}) + (\hat{M}_x^{\prime}, \hat{M}_y^{\prime}, 0)$($\Gamma=\Gamma_6, \Gamma_7$)~\cite{M_comment},}
five electric quadrupoles 
$\rev{(\hat{Q}_u^{\rm (c)}, \hat{Q}_\varv^{\rm (c)}, \hat{Q}_{yz}^{\rm (c)}, \hat{Q}_{zx}^{\rm (c)}, \hat{Q}_{xy}^{\rm (c)})}$ \rev{when $X$ is $(\hat{Q}_u, \hat{Q}_\varv, \hat{Q}_{yz}, \hat{Q}_{zx}, \hat{Q}_{xy})$}, 
\rev{and}
two magnetic octupoles $(\hat{M}_{xyz}^{\rm (c)}, \hat{M}_{z}^{\beta{\rm (c)}})$
\rev{when $X$ is $(\hat{M}_{xyz}, \hat{M}_{z}^{\beta})$}. 
Meanwhile, 
\rev{since the staggered component $\rev{\hat{X}_{\rm A}-\hat{X}_{\rm B}}$ shows odd-parity with respect to the spatial inversion operation,}
the \rev{cluster} odd-parit\rev{y m}ultipoles are 
\rev{defined by} the staggered component 
as electric dipoles $(\hat{Q}_x^{\rm (c)}, \hat{Q}_y^{\rm (c)}, \hat{Q}_z^{\rm (c)})$
\rev{when $X$ is $(\hat{Q}_{zx}, \hat{Q}_{yz}, \hat{Q}_0+\hat{Q}_u)$}~\cite{Q_comment}, 
magnetic toroidal dipoles $(\hat{T}_y^{\rm (c)}, -\hat{T}_x^{\rm (c)})$
\rev{when $X$ is $\sum_{\Gamma} (\hat{M}_x^{\Gamma}, \hat{M}_y^{\Gamma}) + (\hat{M}_x^{\prime}, \hat{M}_y^{\prime})$}~\cite{M_comment}, 
electric toroidal quadrupoles 
$(\hat{G}_{xy}^{\rm (c)}, \hat{G}_{\varv}^{\rm (c)})$ 
\rev{when $X$ is $(\hat{Q}_{\varv}, \hat{Q}_{xy})$}, 
and magnetic quadrupoles 
$(\hat{M}_u^{\rm (c)}, \hat{M}_{xy}^{\rm (c)}, \hat{M}_{\varv}^{\rm (c)})$
\rev{when $X$ is $(\sum_\Gamma\hat{M}_z^{\Gamma}, \hat{M}_{xyz}, \hat{M}_{z}^{\beta})$}~\cite{M_comment}. 
For clarity, we introduce the superscript ``$({\rm c})$" as the notation for cluster multipoles.
The correspondence of local and cluster multipoles is summarized in Table~\ref{table:multi_corres}.
\rev{Hereafter, we use the notations of the cluster multipoles $\hat{X}^{\rm (c)}$ instead of $\hat{X}_{\rm A}+\hat{X}_{\rm B}$ and $\hat{X}_{\rm A}-\hat{X}_{\rm B}$ to clearly represent the effect of the odd-parity multipoles on NQR and NMR spectra.
}

\tabcolsep = 5pt
\begin{table}[htb!]
\centering
\caption{
\rev{
Correspondence of \rev{(a)} uniform and \rev{(b)} staggered components, $
\rev{\hat{X}_{\rm A}+\hat{X}_{\rm B}} $ and $\rev{\hat{X}_{\rm A}-\hat{X}_{\rm B}}$, to \rev{cluster} even-parit\rev{y m}ultipoles (EPMP) and \rev{cluster} odd-parit\rev{y m}ultipoles (OPMP).
Magnetic dipoles $(\hat{M}_x^{\rm tot}, \hat{M}_y^{\rm tot}, \hat{M}_z^{\rm tot})$ represent $\hat{M}^{\rm tot}_\mu = \sum_\Gamma \hat{M}_\mu^{\Gamma}+\hat{M}'_\mu$ for $\mu=x, y$ and $\hat{M}^{\rm tot}_z = \sum_\Gamma \hat{M}_z^{\Gamma}$, where $\Gamma=\Gamma_6, \Gamma_7$~\cite{M_comment}.
\rev{In the notation of the types of multipoles (MP), E, M, ET, MT represent electric, magnetic, electric toroidal, and magnetic toroidal, respectively.}
\label{table:multi_corres}}
}
\rev{
\begin{tabular}{ccl}
\multicolumn{3}{l}{(a) uniform component} \\
\hline \hline
uniform component & EPMP & type of MP\\ \hline
$\hat{Q}_{0,{\rm A}}+\hat{Q}_{0,{\rm B}}$ & $\hat{Q}_0^{\rm (c)}$ & E monopole \\
$\hat{M}_{x,{\rm A}}^{\rm tot}+\hat{M}_{x,{\rm B}}^{\rm tot}$ & $\hat{M}_x^{\rm (c)}$ & $x$-type M dipole \\
$\hat{M}_{y,{\rm A}}^{\rm tot}+\hat{M}_{y,{\rm B}}^{\rm tot}$ & $\hat{M}_y^{\rm (c)}$ & $y$-type M dipole \\
$\hat{M}_{z,{\rm A}}^{\rm tot}+\hat{M}_{z,{\rm B}}^{\rm tot}$ & $\hat{M}_z^{\rm (c)}$ & $z$-type M dipole \\
$\hat{Q}_{u,{\rm A}}+\hat{Q}_{u,{\rm B}}$ & $\hat{Q}_u^{\rm (c)}$ & $3z^2-r^2$-type E quadrupole \\
$\hat{Q}_{\varv,{\rm A}}+\hat{Q}_{\varv,{\rm B}}$ & $\hat{Q}_\varv^{\rm (c)}$ & $x^2-y^2$-type E quadrupole \\
$\hat{Q}_{yz,{\rm A}}+\hat{Q}_{yz,{\rm B}}$ & $\hat{Q}_{yz}^{\rm (c)}$ & $yz$-type E quadrupole \\
$\hat{Q}_{zx,{\rm A}}+\hat{Q}_{zx,{\rm B}}$ & $\hat{Q}_{zx}^{\rm (c)}$ & $zx$-type E quadrupole \\
$\hat{Q}_{xy,{\rm A}}+\hat{Q}_{xy,{\rm B}}$ & $\hat{Q}_{xy}^{\rm (c)}$ & $xy$-type E quadrupole \\
$\hat{M}_{xyz,{\rm A}}+\hat{M}_{xyz,{\rm B}}$ & $\hat{M}_{xyz}^{\rm (c)}$ & $xyz$-type M octupole \\
$\hat{M}_{z,{\rm A}}^\beta+\hat{M}_{z,{\rm B}}^\beta$ & $\hat{M}_{z}^{\beta {\rm (c)}}$ & $z(x^2-y^2)$-type M octupole \\
\hline \hline
\\
\multicolumn{3}{l}{(b) staggered component} \\
\hline \hline
uniform component & OPMP & type of MP\\ \hline
$\hat{Q}_{0,{\rm A}}-\hat{Q}_{0,{\rm B}}$ & $\hat{Q}_z^{\rm (c)}$ & $z$-type E dipole \\
$\hat{M}_{x,{\rm A}}^{\rm tot}-\hat{M}_{x,{\rm B}}^{\rm tot}$ & $\hat{T}_y^{\rm (c)}$ & $y$-type MT dipole \\
$\hat{M}_{y,{\rm A}}^{\rm tot}-\hat{M}_{y,{\rm B}}^{\rm tot}$ & $-\hat{T}_x^{\rm (c)}$ & $x$-type MT dipole \\
$\hat{M}_{z,{\rm A}}^{\rm tot}-\hat{M}_{z,{\rm B}}^{\rm tot}$ & $\hat{M}_u^{\rm (c)}$ & $3z^2-r^2$-type M quadrupole \\
$\hat{Q}_{u,{\rm A}}-\hat{Q}_{u,{\rm B}}$ & $\hat{Q}_z^{\rm (c)}$ & $z$-type E dipole \\
$\hat{Q}_{\varv,{\rm A}}-\hat{Q}_{\varv,{\rm B}}$ & $\hat{G}_{xy}^{\rm (c)}$ & $xy$-type ET quadrupole \\
$\hat{Q}_{yz,{\rm A}}-\hat{Q}_{yz,{\rm B}}$ & $\hat{Q}_{y}^{\rm (c)}$ & $y$-type E dipole \\
$\hat{Q}_{zx,{\rm A}}-\hat{Q}_{zx,{\rm B}}$ & $\hat{Q}_{x}^{\rm (c)}$ & $x$-type E dipole \\
$\hat{Q}_{xy,{\rm A}}-\hat{Q}_{xy,{\rm B}}$ & $\hat{G}_{\varv}^{\rm (c)}$ & $x^2-y^2$-type ET quadrupole \\
$\hat{M}_{xyz,{\rm A}}-\hat{M}_{xyz,{\rm B}}$ & $\hat{M}_{xy}^{\rm (c)}$ & $xy$-type M quadrupole \\
$\hat{M}_{z,{\rm A}}^\beta-\hat{M}_{z,{\rm B}}^\beta$ & $\hat{M}_{\varv}^{{\rm (c)}}$ & $x^2-y^2$-type M quadrupole \\
\hline \hline
\end{tabular}
}
\end{table}

\subsection{Local Hamiltonian for 4$f$ electrons \label{sec:local_Hamiltonian}}
To examine a hyperfine field on $^{59}$Co nucleus, we need to take into account an effective field generated from Ce site. 
As described below, an effective hyperfine field on $^{59}$Co nucleus depends on types of multipole orderings of $4f$ electrons at Ce site. 
We here introduce a local Hamiltonian for Ce electron at the phenomenological level to incorporate the effect of odd-parity multipoles.
The local Hamiltonian for $i$ sublattice is given by 
\begin{align}
\label{eq:local_model}
\mathcal{H}_{{\rm Ce}_i} 
=& \Delta (\hat{Q}_{0i}+\hat{Q}_{ui})
- {\bm H}^{\rm (el)} \cdot \hat{\bm M}_i \mp  \sum_X h^{\rm s}_X \hat{X}_i.
\end{align}
$\Delta$ in the first term is the crystal-field energy of the $\Gamma_6$ level measured from the $\Gamma_7$ level ($\Delta>0$), which is estimated as $\sim100$~K~\cite{tanida2018substitution}.
We set $\Delta=0.5$ in the following calculation.
The second term in Eq.~\eqref{eq:local_model} is the Zeeman term for ${\bm H}^{\rm (el)}\equiv \mu_{\rm B}{\bm H}$ coupled with the magnetic dipoles ${\bm M} = (M_x, M_y, M_z)$, where 
$\mu_{\rm B}$ and ${\bm H}$ are Bohr magneton and magnetic field, respectively. 
We take the linear combination of intraorbital components 
$\hat{M}_\mu^{\Gamma_6}, \hat{M}_\mu^{\Gamma_7}$ and interorbital component $\hat{M}_\mu^\prime$ as 
$\hat{M}_{\mu} \equiv (\hat{M}_\mu^{\Gamma_7} +\delta^{\Gamma_6}\hat{M}_\mu^{\Gamma_6}\pm\delta' \hat{M}'_{\mu})$ 
[the sign is $+(-)$ for $\mu=x(y)$] and 
$\hat{M}_{z} \equiv (\hat{M}_z^{\Gamma_7} +\delta^{\Gamma_6}\hat{M}_{z}^{\Gamma_6})$.
The parameters $\delta^{\Gamma_6}$ and $\delta'$ are introduced to represent the difference of the magnetic susceptibility per different orbitals and are taken to be $(\delta^{\Gamma_6},\delta')=(1/4,1/2)$, which depend on the spin-orbit coupling and the crystal field~\cite{Delta_comment}.
The last term in Eq.~\eqref{eq:local_model} represents the multipolar mean fields leading to the multipole orderings with $\langle \hat{X}_i \rangle \neq 0$\rev{, which mimic the effect of the Coulomb interaction. 
They originate from the mean-field decoupling for the intraorbital and interorbital Coulomb interaction terms~\cite{santini2009multipolar}; the multipoles activated in a $\Gamma_6$ or $\Gamma_7$ level are relevant with the intraorbital Coulomb interaction, while those activated between the $\Gamma_6$ and $\Gamma_7$ levels are relevant with the interorbital Coulomb interaction}. 
As we focus on the cluster multipoles induced by the staggered electronic orderings, we adopt the negative (positive) sign for the A (B) sublattice.
For simplicity, we omit the multipole-multipole interaction between A and B sublattices\rev{, which is renormalized into $h_X^{\rm s}$ at the mean-field level}. 

In the following discussion, we mainly consider three types of staggered orderings: $M_x$-type AFM, $Q_u$-type AFQ, and $Q_\varv$-type AFQ states, whose schematics are shown in Figs.~\ref{fig:figure2}(a)--\ref{fig:figure2}(c), respectively.
This is because the neutron diffraction has indicated the $M_x$-type AFM state at low temperatures~\cite{nikitin2020gradual}. 
On the other hand, as the PIOP is still controversial, we discuss two types of AFQ states for candidates. One is the $Q_u$-type AFQ state arising from the intraorbital multipole degrees of freedom, while the other is the $Q_\varv$-type AFQ state arising from the interorbital multipole degrees of freedom.
For completeness, we also investigate other antiferroic multipole ordered states and the results are summarized in Sec.~\ref{sec:summary_spectra}.

\subsection{Multipoles in AFQ and AFM orderings  \label{sec:local_AF}}

\begin{figure*}[htb!]
\centering
\includegraphics[width=170mm]{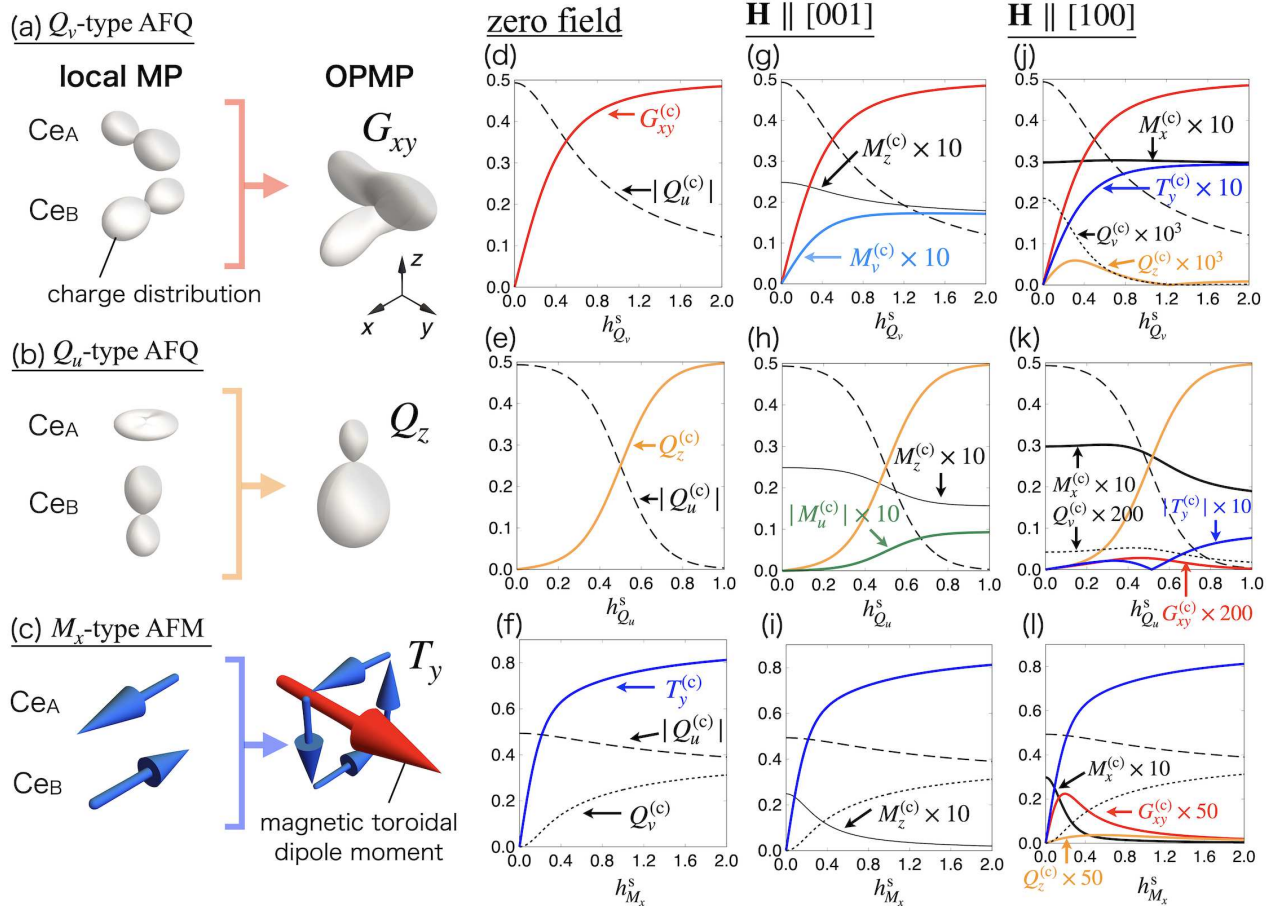}
\caption{
(a)--(c) Schematics of local multipoles (MP) and \rev{cluster} odd-parit\rev{y m}ultipoles (OPMP) in the (a) $Q_\varv$-type AFQ, (b) $Q_u$-type AFQ, and (c) $M_x$-type AFM states.
The shape of the picture in (a) and (b) represents the charge distribution. 
The blue and red arrows in (c) represent the magnetic dipole and magnetic toroidal dipole moments, respectively.
(d)--(l) The staggered mean field dependence of multipoles under (d)--(f) zero magnetic field, (g)--(i) magnetic field ${\bm H} \parallel [001]$ and (j)--(l) ${\bm H} \parallel [100]$.
The data represent those in (d),(g),(j) $Q_\varv$-type AFQ, (e),(h),(k) $Q_u$-type AFQ, and (f),(i),(l) $M_x$-type AFM states, respectively.
Black solid and dashed lines represent the even-parity multipole moments, whereas colored solid lines are odd-parity multipole moments.
\label{fig:figure2}}
\end{figure*}

We discuss the behavior of the electronic multipole moments induced by the staggered mean field with and without the external magnetic field. 
We evaluate the thermal expectation value of the multipole moments  
$X \equiv \braket{\hat{X}} = 
\sum_n \braket{n|\hat{X}|n} \exp{(-\beta E_n)}/Z$, 
where $\ket{n}$ ($n=1$--$8$) is the eigenstate with energy $E_n$ of the total Hamiltonian $\mathcal{H}_{{\rm Ce}_{\rm A}} +\mathcal{H}_{{\rm Ce}_{\rm B}} $, and $Z$ is a partition function. 
We set the inverse temperature $\beta =10$\rev{, which corresponds to $T/\Delta$=0.2}.

Figures~\ref{fig:figure2}(d)--\ref{fig:figure2}(f) show all the nonzero multipole moments at zero magnetic field as a function of the staggered fields $h_{Q_\varv}^{\rm s}$, $h_{Q_u}^{\rm s}$, and $h_{M_x}^{\rm s}$, respectively.
It is noted that $Q_u^{\rm (c)}$ becomes nonzero irrespective of types of order parameters due to nonzero $\Delta$ in Eq.~(\ref{eq:local_model}).
When the mean fields $h_X^{\rm s}$ turn on, the corresponding cluster odd-parity multipole moments $X^{\rm (c)}$ become nonzero.

The results in the $Q_{\varv}$- and $Q_u$-type AFQ ordered states are shown in Figs.~\ref{fig:figure2}(d) and \ref{fig:figure2}(e), respectively.
The odd-parity electric toroidal quadrupole $G^{(\rm c)}_{xy}$ is induced in the $Q_{\varv}$-type AFQ ordering, while the odd-parity electric dipole $Q^{(\rm c)}_z$ is induced in the $Q_u$-type AFQ ordering. 
The mean-field dependence of the odd-parity moments are different with each other: $G^{(\rm c)}_{xy}$ roughly increases as a function of $h^{\rm s}_{Q_{\varv}}$, whereas $Q^{(\rm c)}_{z}$ increases as a function of $(h^{\rm s}_{Q_{u}})^3$ in the small $h_X^{\rm s}$ region. 
This is attributed to the nature of the odd-parity order parameters, which is understood from the perturbation expansion for large $\Delta$, as detailed in Appendix~\ref{ap:power}.

According to the development of $G^{\rm (c)}_{xy}$ or $Q^{\rm (c)}_z$, $Q_u^{\rm (c)}$ is suppressed in both AFQ states in different ways. 
In the case of the $Q_\varv$-type AFQ ordered state,  
$Q_u^{\rm (c)}$ is suppressed as 
$(h_{Q_\varv}^{\rm s})^2$, while it is suppressed as $(h_{Q_u}^{\rm s})^4$ in the $Q_u$-type AFQ state. 
The different mean-field dependences of the multipole moments give different multipole-field dependences of the NQR and NMR frequency shifts, as discussed in Secs.~\ref{sec:NQR} and \ref{sec:NMR}.

Figure~\ref{fig:figure2}(f) shows the result in the $M_x$-type AFM state with the odd-parity magnetic toroidal dipole moment $T_y^{\rm (c)}$. 
The mean-field dependence of $T_y^{\rm (c)}$ is similar to that in the $Q_{\varv}$-type AFQ ordering in Fig.~\ref{fig:figure2}(d).
As a different point, the additional even-parity electric quadrupole $Q_\varv^{\rm (c)}$ is induced in the AFM state, which reflects the breaking of the fourfold rotational symmetry.
In other words, $T_y^{\rm (c)}$ and $Q_\varv^{\rm (c)}$ belong to the same irreducible representation in the AFM state.

Next, we discuss the effect of the magnetic field, whose magnitude is set to be $|\bm{H}^{\rm (el)}|=0.01$.
The results are shown in Figs.~\ref{fig:figure2}(g)--\ref{fig:figure2}(i) in the case of the $[001]$ field and in Figs.~\ref{fig:figure2}(j)--\ref{fig:figure2}(l) in the case of the $[100]$ field. 
There are two important observations under the magnetic field. 
The first one is that additional multipole moments other than the magnetic dipole moments $\bm{M}^{\rm (c)}$ are induced according to the lowering of the crystal symmetry by the magnetic field. 
For example, in the $Q_\varv$-type AFQ state, magnetic quadrupole moment $M_{\varv}^{\rm (c)}$ becomes nonzero for the field along the [001] direction in Fig.~\ref{fig:figure2}(g), while nonzero $Q_{\varv}^{\rm (c)}$, $Q_z^{\rm (c)}$, and  $T_y^{\rm (c)}$ are induced for that along the [100] direction in Fig.~\ref{fig:figure2}(j). 
The second one is that the additional multipole moments induced by the magnetic field are much smaller than primary odd-parity multipole moments, which indicates that the additional multipoles lead to the small quantitative change in the NQR and NMR spectra. 
We summarize the active multipole moments induced by the AFQ and AFM orderings at zero and nonzero fields in Table~\ref{table:finite_multipole}.
The obtained results are consistent with those by the symmetry analysis.

\begin{table}[htb!]
\centering
\caption{
Multipole moments induced in the $Q_\varv$-type AFQ, $Q_u$-type AFQ, and $M_x$-type AFM ordered states as well as the paramagnetic (para) state.
For nonzero fields, additional multipoles induced by $\bm{H}$ are shown. 
\label{table:finite_multipole}}
\begin{tabular}{ccccccccc}\hline \hline
$\bm{H}$ & para & $Q_\varv$-type AFQ & $Q_u$-type AFQ & $M_x$-type AFM \\ \hline
zero & $Q^{\rm (c)}_u$ &  $G_{xy}^{\rm (c)}$ & $Q_z^{\rm (c)}$ & $T_y^{\rm (c)}$, $Q_\varv^{\rm (c)}$  \\
\hline
$\parallel [001]$ & $M_z^{\rm (c)}$ & $M_\varv^{\rm (c)}$ & $M_u^{\rm (c)}$ & ---  \\
$\parallel [100]$ & $M_x^{\rm (c)}$, $Q_\varv^{\rm (c)}$ & $Q_z^{\rm (c)}$, $T_y^{\rm (c)}$ & $G_{xy}^{\rm (c)}$, $T_y^{\rm (c)}$ & $Q_z^{\rm (c)}$, $G_{xy}^{\rm (c)}$\\
\hline \hline
\end{tabular}
\end{table}

\section{Hyperfine field at $\rev{^{59}{\rm Co}}$ nucleus \label{sec:hf}}
We discuss the hyperfine field acting on the nuclear spins at $^{59}$Co ions through effective multipole fields generated in Eq.~(\ref{eq:local_model}). 
In general, the hyperfine field up to the second order of the nuclear spin with $I \geq 1$ is given by~\cite{das1958solid}
\begin{align}
\label{eq:H_nucleus}
\mathcal{H} &= -\gamma \hbar {\bm H}\cdot \hat{\bm I} + \frac{e^2qQ}{4I(2I-1)}\left[3\hat{I}_Z^2-\hat{I}^2+\eta \left(\hat{I}_X^2-\hat{I}_Y^2\right)\right],
\end{align}
where $\hat{\bm{I}}=(\hat{I}_X, \hat{I}_Y, \hat{I}_Z)$ is the nuclear spin operator with the principal axes of the \rev{local} electric-field gradient \rev{at Co nuclear site,} ($X$, $Y$, $Z$).
The magnitude of $\hat{\bm{I}}$ is given by $I=7/2$ for $^{59}$Co nucleus.
The first term represents the Zeeman coupling term; $\gamma$ and $\hbar$ represent gyromagnetic ratio and Dirac's constant, respectively.
The second term describes the nuclear quadrupole interaction; $e$ is the electric charge, $q$ is the electric-field gradient parameter, $Q$ is the nuclear electric quadrupole moment, and $\eta$ is the anisotropic parameter.
The amplitudes of $\bm{H}$, $q$, and $\eta$ depend on electronic multipole moments at neighboring four Ce sites [Fig.~\ref{fig:figure1}(c)] as well as the external magnetic field and crystal-field potential.
When we define ${\bm H}^{\rm (n)} \equiv \gamma \hbar {\bm H}$, the energy scale of the nuclear system is compared with that of the electronic system as ${\bm H}^{\rm (n)}/{\bm H}^{\rm (el)} \sim 10^{-4}$.
We rewrite the Hamiltonian in Eq.~(\ref{eq:H_nucleus}) in terms of the crystal axes coordinate ($x,y,z$) \rev{[see also Fig.~\ref{fig:figure1}(a)]} as 
\begin{align}
\label{eq:H_nucleus_g}
\mathcal{H}=& {\bm C} \cdot \hat{\bm I}
+C_u \hat{I}_u + C_\varv \hat{I}_\varv + C_{yz} \hat{I}_{yz} + C_{zx} \hat{I}_{zx} + C_{xy} \hat{I}_{xy},
\end{align}
where 
\begin{align}
\hat{I}_u&=\frac{1}{2}\left(3\hat{I}_z^2-\hat{I}^2\right), \\
\hat{I}_\varv &= \frac{\sqrt{3}}{2}\left(\hat{I}_x^2-\hat{I}_y^2\right), \\
\hat{I}_{yz}&=\frac{\sqrt{3}}{2}\left(\hat{I}_y\hat{I}_z+\hat{I}_z\hat{I}_y\right), \\
\hat{I}_{zx}&=\frac{\sqrt{3}}{2}\left(\hat{I}_z\hat{I}_x+\hat{I}_x\hat{I}_z\right), \\
\hat{I}_{xy}&=\frac{\sqrt{3}}{2}\left(\hat{I}_x\hat{I}_y+\hat{I}_y\hat{I}_x\right).
\end{align}
The coupling constants for the effective magnetic field and electric-field gradient are parameterized as ${\bm C}=(C_x, C_y, C_z)$ and $(C_u, C_{\varv}, C_{yz}, C_{zx}, C_{xy})$, respectively.
Among them, $C_\mu$($\mu=x, y, z$) includes two contributions from the external field $H_\mu^{\rm (n)}$ and the internal dipole field $C_\mu^{\rm el}$ from the electronic multipoles as 
\begin{align}
\label{eq:C_dipole}
C_\mu&=-{\bm H}^{\rm (n)}+C_{\mu}^{\rm el}, 
\end{align}
whereas $C_{\nu}$ ($\nu = u, \varv, yz,zx,xy$) consists of two contributions from the crystal-field potential $C_{\nu}^{\rm CF}$ and the internal quadrupole field $C_{\nu}^{\rm el}$ from the electronic multipoles as
\begin{align}
\label{eq:C_quadrupole}
C_\nu &= C_{\nu}^{\rm CF} + C_{\nu}^{\rm el}.
\end{align}
In Eqs.~\eqref{eq:C_dipole} and~\eqref{eq:C_quadrupole}, $C_{\mu}^{\rm el}$ and $C_{\nu}^{\rm el}$ depend on types of multipole orderings, which become nonzero through the effective hyperfine coupling between the electronic multipoles and nuclear spins or quadrupoles.

In the following sections, we focus on the multipole contributions to the effective hyperfine field by setting $C_\nu^{\rm CF}=0$ for simplicity~\cite{CCF_comment}. We show an effective Hamiltonian for Co nucleus under multipole fields from Ce sites at zero magnetic field in Sec.~\ref{sec:hf_zero} and at finite magnetic fields in Sec.~\ref{sec:hf_field}.

\subsection{At a zero magnetic field \label{sec:hf_zero}}
Before discussing the effect of odd-parity multipoles, we start from the hyperfine field in the paramagnetic state.
In the paramagnetic state at zero magnetic field, only electric quadrupole $Q_u^{\rm (c)}$ becomes finite among electronic multipoles, which corresponds to the second term in Eq.~\eqref{eq:H_nucleus_g}, as shown in Table~\ref{table:finite_multipole}.
The nuclear Hamiltonian at single Co site is given by
\begin{align}
\label{eq:hf_para}
\mathcal{H}_{\rm para}=C^{\rm el}_u \hat{I}_u
\rev{\equiv c_{u}^{\rm e}Q_u^{\rm (c)} \hat{I}_u, }
\end{align}
where the coupling constant $C_u^{\rm el}$ is represented by the product of the hyperfine coupling constant 
\rev{$c_u^{\rm e}$} and the thermal average of the cluster electronic multipole $Q_u^{\rm (c)}$, $C_u^{\rm el} =
\rev{c_u^{\rm e}}Q_u^{\rm (c)}$.
Here and hereafter, the superscript and subscript in \rev{$c_{\mu}^{\rm p}$}
represent the even- or odd-parity (${\rm p}=$e or o) multipoles and \rev{type of the coupled nuclear multipoles ($\mu=x,y,z,u,\varv,yz,zx,xy$)}, respectively.

The other terms in Eq.~\eqref{eq:H_nucleus_g} become nonzero once the electronic multipole orderings occur, i.e., for nonzero $h^{\rm s}_X$ in Eq.~(\ref{eq:local_model}). 
One can derive the effective hyperfine field in the multipole orderings on the basis of magnetic point group symmetry, as it consists of the coupling terms belonging to the totally symmetric representation under $\bar{4}m21'$.
We display the irreducible representations of the cluster multipoles and nuclear multipoles in Table~\ref{table:irrep_multipole}.

\tabcolsep = 0pt
\begin{table}[htb!]
\centering
\caption{Irreducible representations of nuclear multipoles (NMP) and electronic cluster multipoles (CMP) in the local symmetry of the Co site under zero and nonzero magnetic fields ${\bm H}$.
$X_{\pm} \equiv X_x \pm nX_y$ and $X_{2\pm}\equiv X_{yz}\pm nX_{zx}$ for $X=I, Q^{\rm (c)}, M^{\rm (c)}, T^{\rm (c)}$.
 $n=i (1)$ for $\bar{4}m'2'$ ($2'22'$, $2'$). 
For ${\bm H}_{\parallel [001]}$, the multipoles in the square brackets are also activated.
 The irreducible representations of a magnetic point group are represented by using the irreducible representations of its unitary subgroup, which is shown below the respective magnetic point groups~\cite{cracknell1966corepresentations}. The superscript $\pm$ of the irreducible representation is the parity with respect to the anti-linear-unitary operation (even: $+$, odd: $-$).
 The axes of the twofold rotation $C_2$ of $2'22'$ and $\mathcal{T}C_2$ of $2'$ under ${\bm H}_{\perp[001]}$ (${\bm H}_{\perp[\bar{1}10]}$) are along to the $[110]$ and $[001]$ ($[\bar{1}10]$), respectively. The mirror plane in $m'$ is normal to the $[010]$ direction.
\label{table:irrep_multipole}}
\begin{tabular}{ccccccccc}
\hline \hline
\multicolumn{2}{c}{magnetic field } & - & ${\bm H}_{\parallel[001]}$ & ${\bm H}_{\parallel[100]}$ & ${\bm H}_{\parallel[110]}$ & ${\bm H} _{\perp[001]}$ & ${\bm H}_{\perp[010]}$ & ${\bm H}_{\perp[\bar{1}10]}$ \\\cline{3-9}
 & & $\bar{4}m21'$ & $\bar{4}m'2'$& $2'mm'$ & $2'22'$ & $2'$ & $m'$ & $2'$ \rule[0mm]{0mm}{3mm} \\
NMP & CMP & ($\bar{4}m2$) & ($\bar{4}$) & ($m$) & ($2$) & ($1$) & ($1$) & ($1$)  \\\hline
$I_u$ & $Q_u^{\rm (c)}$, $G_{xy}^{\rm (c)}$ & A$_1^{+}$ & A$^{+}$ & A$^{\prime+}$ & A$^{+}$ & A$^{+}$ & A$^{+}$ & A$^{+}$\\
- & $G_\varv^{\rm (c)}$ & A$_2^{+}$ & A$^{-}$ & A$^{\prime\prime+}$ & B$^{-}$ & A$^{+}$ & A$^{-}$ & A$^{-}$ \\
$I_{xy}$ & $Q_{xy}^{\rm (c)}$ & B$_1^{+}$ & B$^{+}$ & A$^{\prime\prime+}$ & A$^{+}$ & A$^{+}$ & A$^{-}$ & A$^{+}$ \\
$I_\varv$ & $Q_\varv^{\rm (c)}, Q_z^{\rm (c)}$ &B$_2^{+}$ & B$^{-}$ & A$^{\prime+}$ & B$^{-}$ & A$^{+}$ & A$^{+}$ & A$^{-}$ \\
$I_{yz}$ & $Q_{yz}^{\rm (c)}$ & E$^{+}$ & - & A$^{\prime-}$ & - & A$^{-}$ & A$^{-}$ & - \\
$I_{zx}$ & $Q_{zx}^{\rm (c)}$ & & - & A$^{\prime\prime-}$ & - & A$^{-}$ & A$^{+}$ & - \\ 
- & $Q_x^{\rm (c)}$ &  E$^{+}$ & - & A$^{\prime\prime-}$ & - & A$^{-}$ & A$^{+}$ & - \\
- & $Q_y^{\rm (c)}$ & & - & A$^{\prime-}$ & - & A$^{-}$ & A$^{-}$ & -\\
\hline
$I_{2+}$ & $Q_{2+}^{\rm (c)}$ $[iQ_{+}^{\rm (c)}]$ &- & E$^{(2)+}$ & - & B$^{+}$ & A$^{-}$ & - & A$^{+}$\\
$I_{2-}$ & $Q_{2-}^{\rm (c)}$ $[iQ_{-}^{\rm (c)}]$ & - & E$^{(1)+}$ & - & A$^{-}$ & A$^{-}$ & - & A$^{-}$ \\
 $[iI_{2+}]$ & $Q_{+}^{\rm (c)}$ $[iQ_{2+}^{\rm (c)}]$  & - & E$^{(2)-}$ & - & A$^{-}$ & A$^{-}$ & - & A$^{-}$ \\
$[iI_{2-}]$ & $Q_{-}^{\rm (c)}$ $[iQ_{2-}^{\rm (c)}]$  & - & E$^{(1)-}$ & - & B$^{+}$ & A$^{-}$ & - & A$^{+}$ \\\hline \hline
- & $M_{xy}^{\rm (c)}$ &  A$_1^{-}$ & A$^{-}$ & A$^{\prime-}$ & A$^{-}$ & A$^{-}$ & A$^{-}$ & A$^{-}$\\
$I_z$ & $M_z^{\rm (c)}, M_\varv^{\rm (c)}$ &  A$_2^{-}$ & A$^{+}$ & A$^{\prime\prime-}$ & B$^{+}$ & A$^{-}$ & A$^{+}$ & A$^{+}$\\
- & $M_z^{\beta {\rm (c)}}, M_u^{\rm (c)}$ & B$_1^{-}$ & B$^{-}$ & A$^{\prime\prime-}$ & A$^{-}$ & A$^{-}$ & A$^{+}$ & A$^{-}$\\
- & $M_{xyz}^{\rm (c)}$ & B$_2^{-}$ & B$^{+}$ & A$^{\prime-}$ & B$^{+}$ & A$^{-}$ & A$^{-}$ & A$^{+}$\\
$I_x$ & $M_x^{\rm (c)}$ &  E$^{-}$ & - & A$^{\prime+}$ & - & A$^{+}$ & A$^{+}$ & - \\ 
$I_y$ & $M_y^{\rm (c)}$ & & - & A$^{\prime\prime+}$ & - & A$^{+}$ & A$^{-}$ & - \\
- & $T_x^{\rm (c)}$ & E$^{-}$ & - & A$^{\prime\prime+}$ & - & A$^{+}$ & A$^{-}$ & - \\
- & $T_y^{\rm (c)}$ & & - & A$^{\prime+}$ & - & A$^{+}$ & A$^{+}$ & - \\
\hline
$I_{+}$ & $M_{+}^{\rm (c)}$ $[iT_{-}^{\rm (c)}]$ & - & E$^{(1)-}$ & - & A$^{+}$ & A$^{+}$ & - & A$^{+}$ \\
$I_{-}$ & $M_{-}^{\rm (c)}$ $[iT_{+}^{\rm (c)}]$ & - & E$^{(2)-}$ & - & B$^{-}$ & A$^{+}$ & - & A$^{-}$\\
$[iI_{-}]$ & $T_{+}^{\rm (c)}$ $[iM_{-}^{\rm (c)}]$ & - & E$^{(2)+}$ & - & A$^{+}$ & A$^{+}$ & - & A$^{+}$\\
$[iI_{+}]$ & $T_{-}^{\rm (c)}$ $[iM_{+}^{\rm (c)}]$ & -& E$^{(1)+}$ & - & B$^{-}$ & A$^{+}$ & - & A$^{-}$ \\
\hline \hline
\end{tabular}
\end{table}

The general form of the effective hyperfine field in the odd-parity multipole orders is given by
\begin{align}
 \label{eq:hf_order_o}
 \mathcal{H}_{\rm order}^{\rm o} =& \rev{c_{z}^{\rm o}} M_\varv^{\rm (c)} \hat{I}_z + \rev{c_{x,y}^{\rm o}} \left(T_y^{\rm (c)} \hat{I}_x+T_x^{\rm (c)} \hat{I}_y \right) \notag\\
&+ \rev{c_{u}^{\rm o}} G_{xy}^{\rm (c)} \hat{I}_u + \rev{c_{\varv}^{\rm o}} Q_z^{\rm (c)} \hat{I}_\varv 
+ \rev{c_{yz,zx}^{\rm o}}\left(Q_{y}^{\rm (c)} \hat{I}_{yz}-Q_x^{\rm (c)} \hat{I}_{zx}\right), \\
\label{eq:hf_order_e}
\mathcal{H}_{\rm order}^{\rm e}=& \rev{c_{z}^{\rm e}}M_z^{\rm (c)} \hat{I}_z 
 +  \rev{c_{x,y}^{\rm e}} \left(M_x^{\rm (c)} \hat{I}_x+M_y^{\rm (c)} \hat{I}_y\right) \notag\\
&+\rev{c_{xy}^{\rm e}} Q_{xy}^{\rm (c)} \hat{I}_{xy} + \rev{c_{\varv}^{\rm e}} Q_\varv^{\rm (c)} \hat{I}_\varv + \rev{c_{yz,zx}^{\rm e}} \left(Q_{yz}^{\rm (c)} \hat{I}_{yz} + Q_{zx}^{\rm (c)} \hat{I}_{zx} \right), 
\end{align}
where $ \mathcal{H}_{\rm order}^{\rm o}$ ($\mathcal{H}_{\rm order}^{\rm e}$) stands for the hyperfine field in the presence of odd-(even-)parity multipoles.
\rev{Interestingly, the effective hyperfine field includes the coupling between electronic odd-parity multipoles and nuclear even-parity multipoles owing to the lack of the local inversion symmetry at the Co site.}
The hyperfine fields in Eqs.~(\ref{eq:hf_para})--(\ref{eq:hf_order_e}) are summarized in Table~\ref{table:hf_coupling_1}(a).

In CeCoSi, there are two Co ions in the unit cell, which are connected by the fourfold rotation. 
As the sign of the odd-parity crystal field at two Co ions is opposite, while that of the even-parity one is same, the total nuclear Hamiltonian in a unit cell
 is given by
\begin{align}
\label{eq:hf}
\mathcal{H}_{{\rm Co}} =&\mathcal{H}_{{\rm Co}_{\rm A}} +\mathcal{H}_{{\rm Co}_{\rm B}}, \\
\mathcal{H}_{{\rm Co}_{\rm A}} 
=&
\mathcal{H}_{\rm para} + \mathcal{H}_{\rm order}^{\rm o}+ \mathcal{H}_{\rm order}^{\rm e},\\
\mathcal{H}_{{\rm Co}_{\rm B}} 
=&
\mathcal{H}_{\rm para} - \mathcal{H}_{\rm order}^{\rm o}+ \mathcal{H}_{\rm order}^{\rm e}. 
\end{align}
The different sign of $ \mathcal{H}_{\rm order}^{\rm o}$ for the different sublattices is an important outcome of odd-parity multipoles. 
In other words, the presence of the sublattice-dependent splitting of the resonant spectrum corresponds to the emergent odd-parity multipoles within the ${\bm q}={\bm 0}$ orders, as shown in Secs.~\ref{sec:NQR} and \ref{sec:NMR}. 
The obtained hyperfine field including the odd-parity multipole moments in Eq.~\eqref{eq:hf_order_o} is one of the main results in this paper.


\tabcolsep = 3pt
\begin{table*}[htb!]
\centering
\caption{
\rev{(a) Hyperfine field at zero magnetic field.
(b), (c) Additional hyperfine field terms in the magnetic field along the (b) $[001]$ and (c) $[100]$ directions. }
The coupling constants are real. 
\label{table:hf_coupling_1}}
\begin{flushleft}
\begin{tabular}{ccccccccc}
\multicolumn{9}{l}{(a)~zero magnetic field} \\\hline \hline
 & $C_x^{\rm el}$ & $C_y^{\rm el}$ & $C_z^{\rm el}$ & $C_u^{\rm el}$ & $C_\varv^{\rm el}$ & $C_{yz}^{\rm el}$ & $C_{zx}^{\rm el}$ & $C_{xy}^{\rm el}$ \\ \hline
$\mathcal{H}_{\rm para}$ & --- & --- & --- &  $\rev{c_{u}^{\rm e}} Q_u^{\rm (c)}$ & --- & --- & --- & ---  \\ 

$\mathcal{H}_{\rm order}^{\rm o}$ & $\rev{c_{x,y}^{\rm o}} T_y^{\rm (c)}$ & $\rev{c_{x,y}^{\rm o}} T_x^{\rm (c)}$ & $\rev{c_{z}^{\rm o}} M_\varv^{\rm (c)}$ & $\rev{c_{u}^{\rm o}} G_{xy}^{\rm (c)}$ & $\rev{c_{\varv}^{\rm o}} Q_z^{\rm (c)}$ & $\rev{c_{yz,zx}^{\rm o}}Q_{y}^{\rm (c)}$ & $-\rev{c_{yz,zx}^{\rm o}}Q_{x}^{\rm (c)}$ & --- \\

 $\mathcal{H}_{\rm order}^{\rm e}$ & $ \rev{c_{x,y}^{\rm e}} M_x^{\rm (c)}$ & $ \rev{c_{x,y}^{\rm e}} M_y^{\rm (c)}$ & $\rev{c_{z}^{\rm e}}M_z^{\rm (c)}$ & --- & $\rev{c_{\varv}^{\rm e}} Q_\varv^{\rm (c)}$ & $\rev{c_{yz,zx}^{\rm e}} Q_{yz}^{\rm (c)}$ & $\rev{c_{yz,zx}^{\rm e}} Q_{zx}^{\rm (c)}$ & $\rev{c_{xy}^{\rm e}}Q_{xy}^{\rm (c)}$
\\ \hline \hline
\end{tabular}

\begin{tabular}{ccccccccc}
\\
\multicolumn{4}{l}{(b)~$[001]$ magnetic field} \\\hline \hline
  & $C_x^{\rm el}$ & $C_y^{\rm el}$ & $C_z^{\rm el}$  & $C_u^{\rm el}$ & $C_\varv^{\rm el}$ & $C_{yz}^{\rm el}$ & $C_{zx}^{\rm el}$ & $C_{xy}^{\rm el}$  \\ \hline
\rev{$\tilde{\mathcal{H}}_{\rm para}^{[001]}$} & --- & --- & \rev{$\tilde{c}_{z}^{\rm e} Q_u^{\rm (c)}$}  & $\rev{\tilde{c}_{u}^{\rm e} M_z^{\rm (c)}}$ & --- & --- & --- & --- \rule[0mm]{0mm}{3mm}\\

\rev{$\tilde{\mathcal{H}}_{\rm order}^{{\rm o} [001]}$} &\rev{ $\tilde{c}_{x,y}^{\rm o}Q_x^{\rm (c)}$} & \rev{$-\tilde{c}_{x,y}^{\rm o}Q_y^{\rm (c)}$} & 
\rev{$\tilde{c}_{z}^{\rm o}G_{xy}^{\rm (c)}$} & \rev{$\tilde{c}_{u}^{\rm o}M_{\varv}^{\rm (c)}$} & \rev{$\tilde{c}_{\varv}^{\rm o}M_{u}^{\rm (c)}$} & \rev{$\tilde{c}_{yz,zx}^{\rm o}T_{x}^{\rm (c)}$} & \rev{$\tilde{c}_{yz,zx}^{\rm o}T_{y}^{\rm (c)}$} & --- \\

\rev{$\tilde{\mathcal{H}}_{\rm order}^{{\rm e} [001]}$} & \rev{$\tilde{c}_{x,y}^{\rm e}Q_{zx}^{\rm (c)}$} & \rev{$\tilde{c}_{x,y}^{\rm e}Q_{yz}^{\rm (c)}$} & ---  & --- & \rev{$\tilde{c}_{\varv}^{\rm e}M_{z}^{\beta {\rm (c)}}$} & \rev{$\tilde{c}_{yz,zx}^{\rm e}M_{y}^{\rm (c)}$} & \rev{$\tilde{c}_{yz,zx}^{\rm e}M_{x}^{\rm (c)}$} & \rev{$\tilde{c}_{xy}^{\rm e}M_{xyz}^{\rm (c)}$} \\ \hline\hline
\end{tabular}
\begin{tabular}{cccc}
\\
\multicolumn{4}{l}{(c)~$[100]$ magnetic field} \\\hline \hline
 & $C_x^{\rm el}$ & $C_y^{\rm el}$ & $C_z^{\rm el}$  \\ \hline
\rev{$\tilde{\mathcal{H}}_{\rm para}^{[100]}$} &\rev{ $\tilde{c}_{x}^{\rm e,1}Q_u^{\rm (c)}+\tilde{c}_{x}^{\rm e,2}Q_\varv^{\rm (c)}$} & --- & --- \rule[0mm]{0mm}{3mm} \\

\rev{$\tilde{\mathcal{H}}_{\rm order}^{{\rm o}[100]}$} & \rev{$\tilde{c}_{x}^{\rm o,1}Q_{z}^{\rm (c)}+\tilde{c}_{x}^{\rm o,2}G_{xy}^{\rm (c)}$} & \rev{$\tilde{c}_{y}^{\rm o,1}G_{\varv}^{\rm (c)}+\tilde{c}_{y}^{\rm o,2}T_{x}^{\rm (c)}$} & \rev{$\tilde{c}_{z}^{\rm o,1}Q_{x}^{\rm (c)}+\tilde{c}_{z}^{\rm o,2}M_{u}^{\rm (c)}$}  \\

\rev{$\tilde{\mathcal{H}}_{\rm order}^{{\rm e}[100]}$} & --- & \rev{$\tilde{c}_{y}^{\rm e,1}Q_{xy}^{\rm (c)}+\tilde{c}_{y}^{\rm e,2}M_{y}^{\rm (c)}$} & \rev{$\tilde{c}_{z}^{\rm e,1}Q_{zx}^{\rm (c)}+ \tilde{c}_{z}^{\rm e,2}M_{z}^{\beta {\rm (c)}}$} \\
\hline \hline
 \end{tabular}
\begin{tabular}{ccccccccc}
\\
\hline \hline
 & $C_u^{\rm el}$ & $C_\varv^{\rm el}$ & $C_{yz}^{\rm el}$ & $C_{zx}^{\rm el}$ & $C_{xy}^{\rm el}$ \\ \hline
\rev{$\tilde{\mathcal{H}}_{\rm para}^{[100]}$} & \rev{$\tilde{c}_{u}^{\rm e,1}Q_\varv^{\rm (c)}+\tilde{c}_{u}^{\rm e,2}M_x^{\rm (c)}$} &  \rev{$\tilde{c}_{\varv}^{\rm e,1}Q_u^{\rm (c)}+\tilde{c}_{\varv}^{\rm e,2}M_x^{\rm (c)}$} & --- & --- & --- \rule[0mm]{0mm}{3mm}\\
\rev{$\tilde{\mathcal{H}}_{\rm order}^{{\rm o}[100]}$} & \rev{$\tilde{c}_{u}^{\rm o,1}Q_{z}^{\rm (c)}+\tilde{c}_{u}^{\rm o,2}T_{y}^{\rm (c)}$} & \rev{$\tilde{c}_{\varv}^{\rm o,1}G_{xy}^{\rm (c)}+\tilde{c}_{\varv}^{\rm o,2}T_{y}^{\rm (c)}$} & \rev{$\tilde{c}_{yz}^{\rm o,1}Q_{y}^{\rm (c)}+\tilde{c}_{yz}^{\rm o,2}M_{xy}^{\rm (c)}$} & \rev{$ \tilde{c}_{zx}^{\rm o,1}M_{u}^{\rm (c)}+\tilde{c}_{zx}^{\rm o,2}M_{\varv}^{\rm (c)}$} & \rev{$\tilde{c}_{xy}^{\rm o,1}G_{\varv}^{\rm (c)}+\tilde{c}_{xy}^{\rm o,2}T_{x}^{\rm (c)}$} \\

\rev{$\tilde{\mathcal{H}}_{\rm order}^{{\rm e}[100]}$} & --- & --- & \rev{$\tilde{c}_{yz}^{\rm e,1}Q_{yz}^{\rm (c)}+\tilde{c}_{yz}^{\rm e,2}M_{xyz}^{\rm (c)}$} & \rev{$\tilde{c}_{zx}^{\rm e,1}M_{z}^{\rm (c)} + \tilde{c}_{zx}^{\rm e,2}M_{z}^{\beta {\rm (c)}}$} & \rev{$\tilde{c}_{xy}^{\rm e}M_{y}^{\rm (c)}$}
 \\ \hline \hline
\end{tabular}
\end{flushleft}
\end{table*}

\subsection{At a magnetic field \label{sec:hf_field}}
At an external magnetic field, a Zeeman term is taken into account, which is given by
\begin{align}
\mathcal{H}_{\rm Zeeman} = - {\bm H}^{\rm (n)} \cdot \hat{\bm I}.
\end{align}
Although the Zeeman term induces the magnetic dipole contribution, it also induces additional electronic multipole contributions according to the lowering of the symmetry.

By considering the magnetic field along the [001] direction, 
\rev{additional hyperfine field terms appear as follow.}
\rev{
\begin{align}
\tilde{\mathcal{H}}_{\rm para}^{[001]} =& \tilde{c}_{z}^{\rm e}Q_u^{\rm (c)} \hat{I}_z + \tilde{c}_{u}^{\rm e}M_z^{\rm (c)} \hat{I}_u, 
\label{eq:hf_001_para}
\\
\tilde{\mathcal{H}}_{\rm order}^{{\rm o}[001]} =& \tilde{c}_{z}^{\rm o}G_{xy}^{\rm (c)}  \hat{I}_z + \tilde{c}_{x,y}^{\rm o} \left(Q_{x}^{\rm (c)} \hat{I}_x - Q_{y}^{\rm (c)} \hat{I}_y\right) \notag\\
&+ \tilde{c}_{u}^{\rm o}M_\varv^{\rm (c)} \hat{I}_u
+ \tilde{c}_{\varv}^{\rm o}M_u^{\rm (c)} \hat{I}_\varv 
+\tilde{c}_{yz,zx}^{\rm o} \left(T_x^{\rm (c)} \hat{I}_{yz}+T_y^{\rm (c)} \hat{I}_{zx}\right),
\label{eq:hf_001_order_o}
 \\
\tilde{\mathcal{H}}_{\rm order}^{{\rm e}[001]} =& 
\tilde{c}_{x,y}^{\rm e} \left(Q_{zx}^{\rm (c)}\hat{I}_x + Q_{yz}^{\rm (c)}\hat{I}_y \right) \notag\\
&+ \tilde{c}_{xy}^{\rm e}M_{xyz}^{\rm (c)} \hat{I}_{xy}
+ \tilde{c}_{\varv}^{\rm e}M_z^{\beta{\rm (c)}} \hat{I}_\varv 
 + \tilde{c}_{yz,zx}^{\rm e} \left(M_{y}^{\rm (c)} \hat{I}_{yz}+M_{x}^{\rm (c)} \hat{I}_{zx} \right), 
\label{eq:hf_001_order_e}
\end{align}
}
\rev{where $\tilde{\mathcal{H}}_{\rm para}^{[001]}$ is the \rev{additional} hyperfine field induced by the magnetic field
in the paramagnetic state, while $\tilde{H}_{\rm order}^{{\rm o}[001]}$ ($\tilde{H}_{\rm order}^{{\rm e}[001]}$) is the \rev{additional} hyperfine field in the presence of the odd(even)-parity multipole orderings.
}
\rev{$\tilde{c}_\mu^{p}$ ($p={\rm e}$ or ${\rm o}$, $\mu=u,\varv, yz, zx, xy$) is a magnetic-field dependent coupling constant, which vanishes without the magnetic field. 
}

\rev{The appearance of various multipole contributions in Eqs.~\eqref{eq:hf_001_para}--\eqref{eq:hf_001_order_e} is due to the reduction of the local symmetry at Co site $\bar{4}m21' \to \bar{4}m'2'$.}
Reflecting the breaking of the time-reversal symmetry, the effective couplings between electronic and nuclear multipoles with opposite time-reversal parity appear.
In other words, the electric (magnetic) multipole at Ce site is coupled with the nuclear dipole (quadrupole) at Co site. 
\rev{From the microscopic viewpoint, such a coupling originates from the magnetic multipoles \rev{with spatially anisotropic distributions}, such as magnetic octupole\rev{, }
 which \rev{are described by the coupling between the anisotropic charge distribution and magnetic moment}~\rev{\cite{sakai1997new, sakai1999antiferro}}. 
For instance, in the case of the $Q_{\varv}$-type ordering under the magnetic field along the [001] direction, the magnetic quadrupole $M^{\rm (c)}_{\varv}$ with time-reversal odd is induced as shown in Fig.~\ref{fig:figure2}(g). 
Since $M^{\rm (c)}_{\varv}$ belongs to the same irreducible representation A$^+$ as $I_u$ with time-reversal even under the magnetic point group $\bar{4}m'2'$ 
from Table~\ref{table:irrep_multipole}, the field-induced $M_{\varv}^{\rm (c)}$ affects the $3z^2-r^2$-type charge distribution and results in the effective coupling between $M_{\varv}^{\rm (c)}$ and $I_u$.
}

Similarly, the \rev{additional} hyperfine fields  
\rev{in} the $[100]$ magnetic field are given by
\rev{
\begin{widetext}
\begin{align}
\tilde{\mathcal{H}}_{\rm para}^{[100]} =& \left(\tilde{c}_{x}^{\rm e,1}Q_u^{\rm (c)} +\tilde{c}_{x}^{\rm e,2}Q_\varv^{\rm (c)} \right)\hat{I}_x
 + \left(\tilde{c}_{u}^{\rm e,1}Q_\varv^{\rm (c)} +\tilde{c}_{u}^{\rm e,2}M_x^{\rm (c)} \right)\hat{I}_u 
 + \left(\tilde{c}_{\varv}^{\rm e,1}Q_u^{\rm (c)}+\tilde{c}_{\varv}^{\rm e,2}M_x^{\rm (c)} \right)\hat{I}_{\varv}, 
\label{eq:hf_100_para}
\\
\tilde{\mathcal{H}}_{\rm order}^{{\rm o}[100]} =&
\left( \tilde{c}_{x}^{\rm o,1}Q_z^{\rm (c)} + \tilde{c}_{x}^{\rm o,2}G_{xy}^{\rm (c)}\right) \hat{I}_x 
+\left(\tilde{c}_{y}^{\rm o,1}G_\varv^{\rm (c)} + \tilde{c}_{y}^{\rm o,2}T_x^{\rm (c)} \right)\hat{I}_y 
+\left( \tilde{c}_{z}^{\rm o,1}Q_x^{\rm (c)} + \tilde{c}_{z}^{\rm o,2}M_u^{\rm (c)} \right)\hat{I}_z  
+\left(\tilde{c}_{u}^{\rm o,1}Q_z^{\rm (c)} + \tilde{c}_{u}^{\rm o,2}T_y^{\rm (c)}\right) \hat{I}_u 
+\left(\tilde{c}_{\varv}^{\rm o,1}G_{xy}^{\rm (c)} + \tilde{c}_{\varv}^{\rm o,2}T_y^{\rm (c)}\right) \hat{I}_\varv 
\notag\\
&+\left(\tilde{c}_{yz}^{\rm o,1}Q_y^{\rm (c)} + \tilde{c}_{yz}^{\rm o,2}M_{xy}^{\rm (c)}\right) \hat{I}_{yz} 
+ \left( \tilde{c}_{zx}^{\rm o,1}M_u^{\rm (c)} + \tilde{c}_{zx}^{\rm o,2}M_\varv^{\rm (c)}\right) \hat{I}_{zx}
+\left(\tilde{c}_{xy}^{\rm o,1}G_\varv^{\rm (c)} + \tilde{c}_{xy}^{\rm o,2}T_x^{\rm (c)}\right) \hat{I}_{xy}, 
\label{eq:hf_100_order_o}
\\
\tilde{\mathcal{H}}_{\rm order}^{{\rm e}[100]} =& 
\left( \tilde{c}_{y}^{\rm e,1}Q_{xy}^{\rm (c)} + \tilde{c}_{y}^{\rm e,2}M_{y}^{\rm (c)}\right) \hat{I}_y
+\left(\tilde{c}_{z}^{\rm e,1}Q_{zx}^{\rm (c)} + \tilde{c}_{z}^{\rm e,2}M_{z}^{\beta {\rm (c)}} \right) \hat{I}_z 
+ \left(\tilde{c}_{yz}^{\rm e,1}Q_{yz}^{\rm (c)} + \tilde{c}_{yz}^{\rm e,2}M_{xyz}^{\rm (c)} \right) \hat{I}_{yz} 
+ \left( \tilde{c}_{zx}^{\rm e,1}M_{z}^{\rm (c)} + \tilde{c}_{zx}^{\rm e,2}M_{z}^{\beta {\rm (c)}} \right) \hat{I}_{zx} 
+ \tilde{c}_{xy}^{\rm e}M_{y}^{\rm (c)}  \hat{I}_{xy}, 
\label{eq:hf_100_order_e}
\end{align}
\end{widetext}
}
where the local symmetry at Co site reduces as $\bar{4}m21' \to 2'mm'$.
For in-plane fields, the $\hat{I}_\varv$ term additionally contributes to 
\rev{$\tilde{\mathcal{H}}_{\rm para}^{[100]}$}
 due to the breaking of the fourfold improper rotational symmetry.

The \rev{additional} hyperfine field Hamiltonian at the external magnetic field is summarized in Tables~\ref{table:hf_coupling_1}(b) and ~\ref{table:hf_coupling_1}(c). 
One can obtain the hyperfine field Hamiltonian for other field directions by using the irreducible representation in Table~\ref{table:irrep_multipole}.

In the end, the total Hamiltonian in a unit cell under the magnetic field is given by
\begin{align}
\label{eq:hf_field}
\mathcal{H}_{{\rm Co}} =&\mathcal{H}_{{\rm Co}_{\rm A}} +\mathcal{H}_{{\rm Co}_{\rm B}} \rev{+\tilde{\mathcal{H}}_{{\rm Co}_{\rm A}} +\tilde{\mathcal{H}}_{{\rm Co}_{\rm B}}}, \\
\mathcal{H}_{{\rm Co}_{\rm A}} 
=&
\mathcal{H}_{\rm Zeeman} +  
\mathcal{H}_{\rm para}  + \mathcal{H}_{\rm order}^{\rm o}+ \mathcal{H}_{\rm order}^{\rm e},\\
\mathcal{H}_{{\rm Co}_{\rm B}} 
=&
\mathcal{H}_{\rm Zeeman} + 
\mathcal{H}_{\rm para} - \mathcal{H}_{\rm order}^{\rm o}+ \mathcal{H}_{\rm order}^{\rm e}\rev{,} 
 \\
\rev{\tilde{\mathcal{H}}_{{\rm Co}_{\rm A}} 
=}&
\rev{\tilde{\mathcal{H}}_{\rm para} + \tilde{\mathcal{H}}_{\rm order}^{\rm o}+ \tilde{\mathcal{H}}_{\rm order}^{\rm e},}\\
\rev{\tilde{\mathcal{H}}_{{\rm Co}_{\rm B}}
=}&
\rev{\tilde{\mathcal{H}}_{\rm para} - \tilde{\mathcal{H}}_{\rm order}^{\rm o} + \tilde{\mathcal{H}}_{\rm order}^{\rm e}.} 
\end{align}
We use above nuclear Hamiltonian \rev{$\mathcal{H}_{\rm Co}$} to examine the NMR spectra in the odd-parity multipole orderings in the following sections.

\section{NQR spectra at zero field \label{sec:NQR}}
We examine how odd-parity multipole moments affect an NQR spectrum.
In the paramagnetic state, the nuclear Hamiltonian given by Eq.~(\ref{eq:hf}) leads to three NQR frequencies, $f=\nu_{\rm Q}$, $2\nu_{\rm Q}$, and $3\nu_{\rm Q}$, where $\hbar\nu_{\rm Q} = 3 \rev{c_u^{\rm e}}Q_u^{\rm (c)}$. 
We take $\nu_{\rm Q}=1$ as the frequency unit.

In the following, we show the resonance frequencies in odd-parity multipole orderings in Secs.~\ref{sec:NQR_Qv}--\ref{sec:NQR_Mx}: the $Q_\varv$-type AFQ state with $G^{\rm (c)}_{xy}$ in Sec.~\ref{sec:NQR_Qv}, the $Q_u$-type AFQ state with $Q^{\rm (c)}_z$ in Sec.~\ref{sec:NQR_Qu}, and the $M_x$-type AFM state with $T^{\rm (c)}_y$ in Sec.~\ref{sec:NQR_Mx}.
In the calculations, we set the coupling constant 
in Eqs.~\eqref{eq:hf_para} and \eqref{eq:hf_order_o} as 
\rev{$c_u^{\rm e}=c_{\rm Q}$,}
which \rev{is} 
estimated from the NQR frequenc\rev{y 
 i}n Ref.~\onlinecite{manago2019} 
\rev{as} $c_{\rm Q}=0.13$  
when setting $\gamma\hbar=1$, 
while the 
\rev{coupling constants are set to be $c$ for the primary-induced multipoles and}
to be 
\rev{$c'$} \rev{for the secondary-induce multipoles} as the unknown model parameters for simplicity.

\subsection{Staggered $Q_{\varv}$-type AFQ \label{sec:NQR_Qv}}
We discuss the NQR spectrum in the staggered $Q_\varv$-type AFQ state, where the effective nuclear Hamiltonian is represented by \rev{considering the finite electronic multipoles in Eqs.~\eqref{eq:hf_para}--\eqref{eq:hf_order_e} as}
\begin{align}
\label{eq:hf_NQR_Qv}
\mathcal{H}_{\rm Co_{\rm A/B}} &= \left( c_{\rm Q} Q_u^{\rm (c)} \pm c G_{xy}^{\rm (c)} \right) \hat{I}_u.
\end{align}
The positive (negative) sign in the second term corresponds to $\mathcal{H}_{\rm Co_{\rm A}}$ ($\mathcal{H}_{\rm Co_{\rm B}}$).

\begin{figure}[htb!]
\centering
\includegraphics[width=88mm]{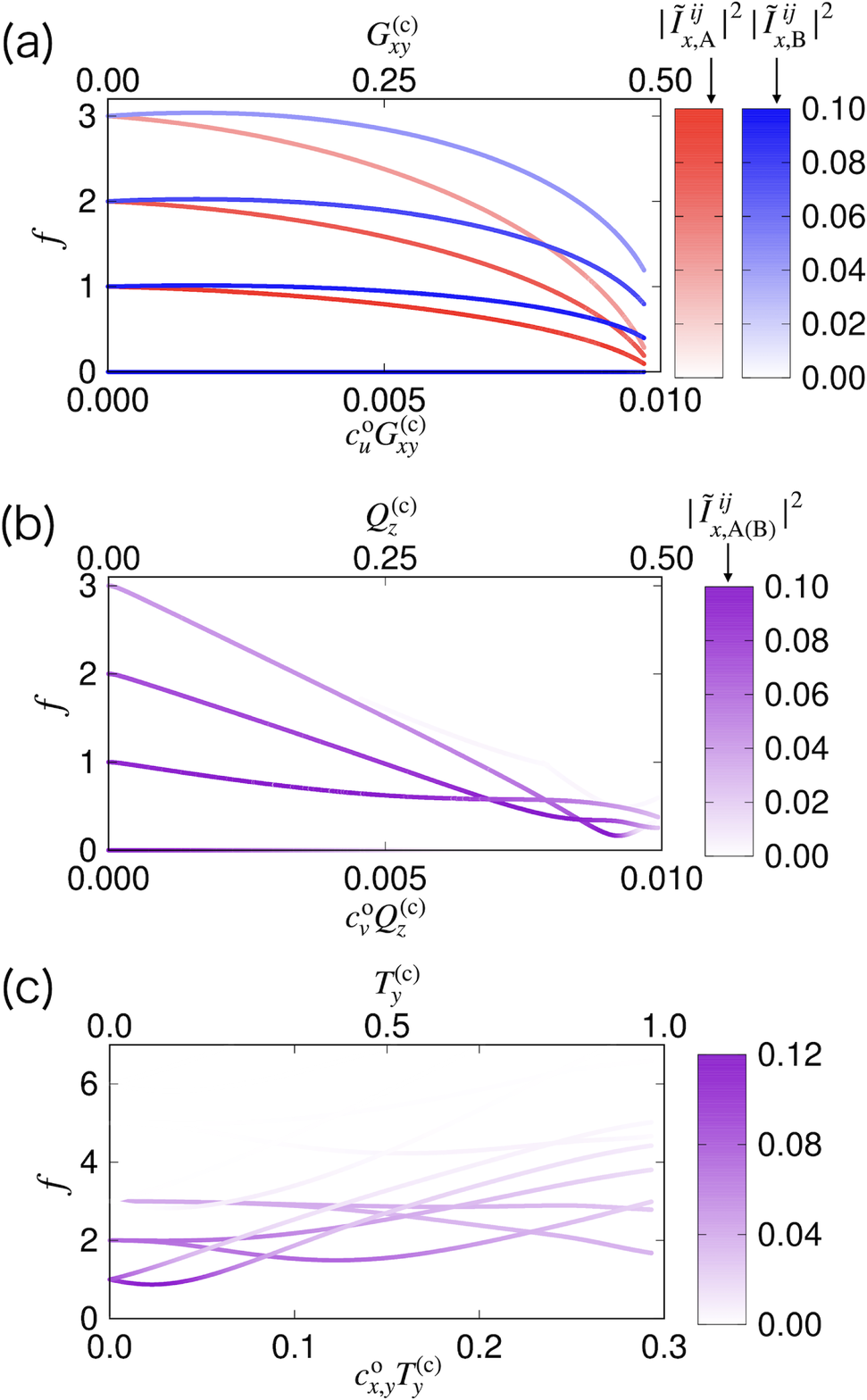}
\caption{The odd-parity multipole (upper scale) and its hyperfine field (lower scale) dependences of the NQR frequency $f$ in the staggered (a) $Q_{\varv}$-type AFQ, (b) $Q_u$-type AFQ, and (c) $M_x$-type AFM states.
The coupling constants \rev{$c_u^{\rm o}$, $c_\varv^{\rm o}$, and $c_{x,y}^{\rm o}$}
are set as \rev{$c_u^{\rm o} = c_\varv^{\rm o} = c = 0.02$ in the AFQ states and $c_{x,y}^{\rm o} = c = 0.3$ in the AFM state.}
\rev{Other coupling constants are set to be $c'=0.02$.}
As an intensity of the spectrum, $\left|\tilde{I}_{x, {\rm A(B)}}^{ij}\right|^2$ is shown by the counter plot in red (blue) for Co$_{\rm A}$ (Co$_{\rm B}$) site.
When the spectra from Co$_{\rm A}$ and Co$_{\rm B}$ are equivalent, their intensities are shown by violet.
\label{fig:figure3}}
\end{figure}

The NQR frequencies of Co$_{\rm A}$ and Co$_{\rm B}$ sites as a function of $G_{xy}^{\rm (c)}$ with fixed $c=0.02$ are shown in Fig.~\ref{fig:figure3}(a).
The color scale in Fig.~\ref{fig:figure3} shows the intensity of the NQR spectrum, which is calculated by the magnitude of the matrix element of $I_x$ between different nuclear state $i$ and $j$ at Co$_{\rm A(B)}$ site, $\left|\tilde{I}_{x,{\rm A(B)}}^{ij}\right|^2 \equiv \left|\braket{i|\tilde{I}_{x, {\rm A(B)}}|j}\right|^2$, where $\tilde{I}_\mu$ ($\mu=x,y,z$) represents the normalized $I_\mu$ satisfying ${\rm Tr}[\tilde{I}_\mu \tilde{I}_\mu^\dagger]=1$.

The result shows that the NQR frequencies for Co$_{\rm A}$ and Co$_{\rm B}$  have different values and show the spectral splittings and shift in the $Q_\varv$-type AFQ state.
The sublattice-dependent splitting is owing to the effective coupling between $G_{xy}^{\rm (c)}$ and $I_u$ with different signs for different sublattices.
In other words, the odd-parity multipole moment $G_{xy}^{\rm (c)}$ in Eq.~(\ref{eq:hf_NQR_Qv}) plays a significant role in splitting of the NQR frequencies.
In fact, the splittings of the NQR frequencies are proportional to $G_{xy}^{\rm (c)}$.
On the other hand, the shift of the frequency to smaller $f$ is due to \rev{the decrease of dominant $c_{\rm Q}Q_u^{\rm (c)}$ \rev{($c_{\rm Q}\gg c$)} term in Eq.~(\ref{eq:hf_NQR_Qv}) by} the suppression of $Q_u^{\rm (c)}$ \rev{while increasing $G_{xy}^{\rm (c)}$ as shown in Fig.~\ref{fig:figure2}(d)}.

Note that it might be difficult to detect the splitting due to the odd-parity multipoles even for a saturated multipole moment $G^{\rm (c)}_{xy} \sim 0.5$ when the coupling constant $c$ is small, since the splittings are proportional to $c G_{xy}^{\rm (c)}$.

\subsection{Staggered $Q_u$-type AFQ \label{sec:NQR_Qu}}
In the staggered $Q_u$-type AFQ state with $Q_z^{\rm (c)}$, the effective nuclear Hamiltonians of Co$_{\rm A}$ and Co$_{\rm B}$ are represented by
\begin{align}
\label{eq:hf_NQR_Qu}
\mathcal{H}_{\rm Co_{\rm A/B}} &=
c_{\rm Q} Q_u^{\rm (c)} \hat{I}_u \pm  c Q_z^{\rm (c)} \hat{I}_\varv.
\end{align}

The NQR spectrum for the coupling constant $c=0.02$ is shown in Fig.~\ref{fig:figure3}(b). 
In contrast to the result in the $Q_{\varv}$-type AFQ state, there is no splitting in the NQR spectrum.
This is because the different sign of $Q_z^{\rm (c)}$ in Eq.~\eqref{eq:hf_NQR_Qu} is not relevant to the splitting, which is consistent with the symmetry argument that there is no linear coupling between $Q_z^{\rm (c)}$ and $Q_u^{\rm (c)}$ in the free energy expansion at Co site.
In the end, nonzero $Q_z^{\rm (c)}$ just affects the spectral shift.

In addition to the splitting, the difference is found in the odd-parity multipole dependence of the frequency shift.
The frequencies in the $Q_u$-type AFQ state in Fig.~\ref{fig:figure3}(b) decrease with increasing $Q_z^{\rm (c)}$ faster than those in the $Q_\varv$-type AFQ state in Fig.~\ref{fig:figure3}(a). 
This is understood from the different dependences on the multipole moments as discussed in Sec.~\ref{sec:local_AF};
$Q_u^{\rm (c)}$ in the $Q_u$-type AFQ state decreases by $\sim [Q_z^{\rm (c)}]^{4/3}$,
 while that in the $Q_\varv$-type AFQ state decreases by $\sim [G_{xy}^{\rm (c)}]^2$.

\subsection{Staggered $M_x$-type AFM \label{sec:NQR_Mx}}
In the staggered $M_x$-type AFM state, the nuclear Hamiltonian is represented by
\begin{align}
\label{eq:NQR_Mx}
\mathcal{H}_{\rm Co_{\rm A/B}} &=
\pm  \rev{c} T_y^{\rm (c)} \hat{I}_x  + c_{\rm Q} Q_u^{\rm (c)} \hat{I}_u + \rev{c'} Q_\varv^{\rm (c)} \hat{I}_\varv. 
\end{align}
It is noted that nuclear dipole contribution in the $M_x$-type AFM appears even without the net magnetization nor the magnetic field.

Figure~\ref{fig:figure3}(c) shows the NQR spectrum for the coupling constant \rev{$c=0.3$ and $c'=0.02$} in the $M_x$-type AFM state\rev{, where $c$ is estimated from the magnitude of the internal magnetic field in Ref.~\onlinecite{manago2019}}. 
The NQR frequencies are split into seven due to the contribution from the internal magnetic field arising from the first term in Eq.~\eqref{eq:NQR_Mx}.
Meanwhile, the NQR frequencies for Co$_{\rm A}$ and Co$_{\rm B}$ sites are the same, which indicates that
there is no sublattice-dependent splitting in the presence of the odd-parity $T_y^{\rm (c)}$. 
This means that $T_y^{\rm (c)}$ does not linearly couple with $Q_u^{\rm (c)}$ in the free energy expansion, which is consistent with the symmetry argument.
Thus, it is difficult to conclude the presence of $T_y^{\rm (c)}$ only from the seven splittings in Fig.~\ref{fig:figure3}(c). 
In fact, the NQR spectra split into seven can be obtained in the even-parity magnetic dipole order, such as $M_x^{\rm (c)}$,  in Table~\ref{table:multi_corres}.

\section{NMR spectra \label{sec:NMR}}
In this section, we discuss the NMR spectra in the odd-parity multipole orderings.
The applied resonance fields are along the $[001]$ and $[100]$ directions in Secs.~\ref{sec:NMR_001} and \ref{sec:NMR_100}, respectively.
We set $\gamma\hbar=1$ and $|{\bm H}^{\rm (n)}|=1$.
The coupling constants are set as 
\rev{$c_u^{\rm e} = c_{\rm Q} = 0.13$}
as well as 
\rev{that} in NQR in Sec.~\ref{sec:NQR}.
\rev{The \rev{other} coupling constants are set to be $c$ \rev{for the primary-induced multipoles} and}
to be \rev{$c'$} \rev{for the secondary-induced multipoles for simplicity}.
The field-swept spectra are shown in Appendix~\ref{ap:NMR_H}.

\subsection{$[001]$-field spectrum \label{sec:NMR_001}}
We discuss the NMR spectra in the paramagnetic state, $Q_\varv$-type AFQ state, $Q_u$-type AFQ state, and $M_x$-type AFM state in Secs.~\ref{sec:NMR_001_para}--\ref{sec:NMR_001_Mx}, respectively.

\subsubsection{Paramagnetic state \label{sec:NMR_001_para}}
In the paramagnetic state at the $[001]$ magnetic field, ${\bm H}^{\rm (n)} = (0,0,H_z^{\rm (n)})$,
the effective nuclear Hamiltonian is represented by
\begin{align}
\label{eq:hf_NMR_001_para1}
\mathcal{H}_{\rm Co_{A/B}} &=
 \rev{\left( - H_z^{\rm (n)} +c' M_z^{\rm (c)}\right) \hat{I}_z
 + c_{\rm Q}Q_u^{\rm (c)}  \hat{I}_u,} \\
 \label{eq:hf_NMR_001_para2}
 \rev{\tilde{\mathcal{H}}_{\rm Co_{A/B}}} & \rev{= c'Q_u^{\rm (c)} \hat{I}_z +c'M_z^{\rm (c)} \hat{I}_u.}
\end{align}
The first term \rev{in Eq.~\eqref{eq:hf_NMR_001_para1}} includes the Zeeman term from the external magnetic field.
The sum of the external magnetic field and the hyperfine field in Eq\rev{s}.~\eqref{eq:hf_NMR_001_para1} \rev{and \eqref{eq:hf_NMR_001_para2}} results in the seven spectral peaks separated by the same interval in the NMR measurement.

\subsubsection{Staggered $Q_\varv$-type AFQ \label{sec:NMR_001_Qv}}
In the $Q_\varv$-type AFQ state, the effective nuclear Hamiltonian is obtained as
\begin{align}
\label{eq:hf_NMR_001_Qv}
\mathcal{H}_{\rm Co_{A/B}} =& 
\rev{\left(-H_z^{\rm (n)}  + c'  M_z^{\rm (c)} \right) \hat{I}_z + \left( c_{\rm Q} Q_u^{\rm (c)} \pm cG_{xy}^{\rm (c)} \right) \hat{I}_u, }\\
 \rev{\tilde{\mathcal{H}}_{\rm Co_{A/B}} =}& 
 \rev{c' \left( Q_u^{\rm (c)} \pm G_{xy}^{\rm (c)} \pm M_\varv^{\rm (c)}
\right)\hat{I}_z
+  c'\left( M_z^{\rm (c)} \pm M_\varv^{\rm (c)}\right) \hat{I}_u.}
\end{align}

\begin{figure}[htb!]
\centering
\includegraphics[width=88mm]{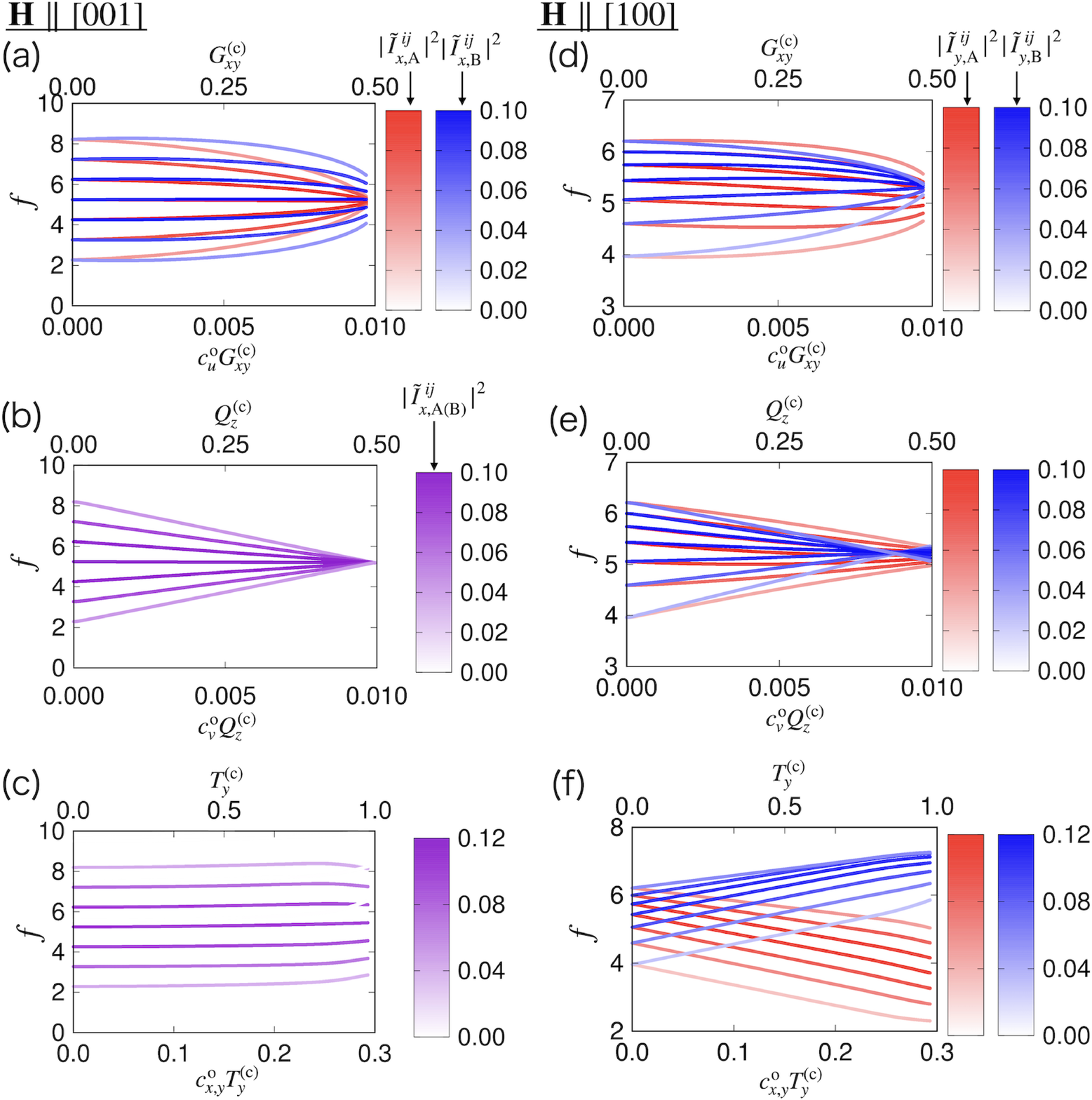}
\caption{The odd-parity multipole dependences of the NMR frequency $f$ under (a)--(c) the $[001]$ magnetic field and (d)--(f) the $[100]$ magnetic field.
The data are for the (a),(d) $Q_{\varv}$-type AFQ, (b), (e) $Q_u$-type AFQ, and (c), (f) $M_x$-type AFM states.
The color scales represent the intensities with (a)--(c) $\left|\tilde{I}_{x, {\rm A(B)}}^{ij}\right|^2$ and (d)--(f) $\left|\tilde{I}_{y, {\rm A(B)}}^{ij}\right|^2$. 
The coupling constants are set as 
\rev{$c_{u}^{\rm o} = c_{\varv}^{\rm o}=c =0.02$ in the AFQ states and $c_{x,y}^{\rm o}=c=0.3$ in the AFM state.}
\rev{Other coupling constants are set to be $c'=0.02$.}
\label{fig:figure4}}
\end{figure}

The frequency-swept NMR spectrum for $c\rev{= c'}=0.02$ is shown in Fig.~\ref{fig:figure4}(a), where the color scale represents the intensity of the $[001]$-field NMR spectrum.
Figure~\ref{fig:figure4}(a) shows that $G_{xy}^{\rm (c)}$ leads to sublattice-dependent spectral splittings due to the different frequencies of Co$_{\rm A}$ and Co$_{\rm B}$ as well as the result in NQR.
\rev{The NMR spectrum is mainly determined by the following dominant contributions: Zeeman term, $c_{\rm Q}Q_u^{\rm (c)}$ term, and primarily induced $G_{xy}^{\rm (c)}$ terms.}
The spectral splittings originate from the odd-parity multipoles $G_{xy}^{\rm (c)}$ and $M_\varv^{\rm (c)}$ which are coupled with $Q_u^{\rm (c)}$ and $M_z^{\rm (c)}$, though the contribution from $M_\varv^{\rm (c)}$ is much smaller than that of $G_{xy}^{\rm (c)}$, as discussed in Sec.~\ref{sec:local_AF}.
Additionally, each spectrum is shifted by $[G_{xy}^{\rm (c)}]^2$ as discussed in  Sec.~\ref{sec:NQR_Qu}.

\subsubsection{Staggered $Q_u$-type AFQ \label{sec:NMR_001_Qu}}
In the $Q_u$-type AFQ state, the effective nuclear Hamiltonian is described as
\begin{align}
\label{eq:hf_NMR_001_Qu}
\mathcal{H}_{\rm Co_{A/B}}
=& 
\rev{\left(-H_z^{\rm (n)} + c' M_z^{\rm (c)} \right) \hat{I}_z + c_{\rm Q} Q_u^{\rm (c)} \hat{I}_u\pm c Q_z^{\rm (c)}  \hat{I}_\varv,} \\
\rev{\tilde{\mathcal{H}}_{\rm Co_{A/B}}
=}& 
\rev{ c' Q_u^{\rm (c)} \hat{I}_z 
+ c' M_z^{\rm (c)}  \hat{I}_u\pm c' M_u^{\rm (c)} \hat{I}_\varv.}
\end{align}

The NMR spectrum for \rev{$c=c'=0.02$} is shown in Fig.~\ref{fig:figure4}(b).
The seven frequencies have no additional split for both Co sites, since the induced odd-parity multipoles, $Q_z^{\rm (c)}$ and $M_u^{\rm (c)}$, in the ordered state do not couple with $Q_u^{\rm (c)}$ or $M_z^{\rm (c)}$. 
Meanwhile, each frequency is shifted by $[Q_z^{\rm (c)}]^{4/3}$, which is understood by the behavior of $Q_u^{\rm (c)}$, as discussed in Sec.~\ref{sec:NQR_Qu}.

For full-saturated $Q_{z}^{\rm (c)}=0.5$, all the NMR frequencies become $f \sim 5.2$, which corresponds to 
the frequency only in the external magnetic field.
This is because $Q_u^{\rm (c)}$ in the 
\rev{crystal-field} term vanishes for $Q_{z}^{\rm (c)}=0.5$, as shown in Fig.~\ref{fig:figure2}(h).

\subsubsection{Staggered $M_x$-type AFM  \label{sec:NMR_001_Mx}}
In the $M_x$-type AFM state, the effective nuclear Hamiltonian for the Co nucleus is represented as
\begin{align}
\label{eq:hf_NMR_001_Mx}
\mathcal{H}_{\rm Co_{A/B}} =& 
\rev{\left(-H_z^{\rm (n)} + c' M_z^{\rm (c)} \right) \hat{I}_z \pm c' T_y^{\rm (c)}\hat{I}_x
+ c_{\rm Q}Q_u^{\rm (c)} \hat{I}_u +c Q_\varv^{\rm (c)} \hat{I}_\varv,} \\
\rev{\tilde{\mathcal{H}}_{\rm Co_{A/B}} =}& 
\rev{c' Q_u^{\rm (c)}  \hat{I}_z 
+ c' M_z^{\rm (c)} \hat{I}_u  \pm c'T_y^{\rm (c)} \hat{I}_{zx}.}
\end{align}

The NMR spectra for \rev{$c=0.3$ and $c'=0.02$} is shown in Fig.~\ref{fig:figure4}(c).
The spectra show no sublattice-dependent splitting, which is similar to those in NQR spectra in Sec.~\ref{sec:NQR_Mx}, as $T_y^{\rm (c)}$ does not couple with $Q_u^{\rm (c)}$ or $M_z^{\rm (c)}$.
The shift of the resonance frequency against $T_y^{\rm (c)}$ is small compared to that in the $Q_u$-type AFQ state in Fig.~\ref{fig:figure4}(b), which reflects the different behavior of $Q_u^{\rm (c)}$, a\rev{s s}hown in Fig.~\ref{fig:figure2}(i). 

\subsection{$[100]$-field spectrum \label{sec:NMR_100}}
We show the $[100]$-field NMR spectrum in the paramagnetic state, $Q_\varv$-type AFQ state, $Q_u$-type AFQ state, and $M_x$-type AFM state in Secs.~\ref{sec:NMR_100_para}--\ref{sec:NMR_100_Mx}, respectively. 

\subsubsection{Paramagnetic state \label{sec:NMR_100_para}}
In the paramagnetic state at the $[100]$ magnetic field,
the effective nuclear Hamiltonian at Co nucleus is represented by 
\begin{align}
\label{eq:NMR_100_para1}
\mathcal{H}_{\rm Co_{A/B}}
=& 
\rev{\left( -H_x^{\rm (n)} + c' M_x^{\rm (c)} \right)\hat{I}_x
+c_{\rm Q} Q_u^{\rm (c)} \hat{I}_u
+c' Q_\varv^{\rm (c)}\hat{I}_{\varv}.} \\
\label{eq:NMR_100_para2}
\rev{\tilde{\mathcal{H}}_{\rm Co_{A/B}}}
=& \rev{
c'\left( Q_u^{\rm (c)} + Q_\varv^{\rm (c)}\right) \hat{I}_x} \notag\\
&\rev{
+c'\left( Q_\varv^{\rm (c)}+M_x^{\rm (c)} \right) \hat{I}_u
+c' \left(Q_u^{\rm (c)} +M_x^{\rm (c)}\right)\hat{I}_{\varv}.}
\end{align}

The nuclear Hamiltonian in Eq\rev{s}.~\eqref{eq:NMR_100_para1} \rev{and \eqref{eq:NMR_100_para2}} 
leads to the seven spectra similar to those at the $[001]$ magnetic field.
However, the intervals between the resonance frequencies are not equivalent, since the magnetic field normal to the $z$ axis leads to the emergence of $Q^{\rm (c)}_\varv$.

\subsubsection{Staggered $Q_\varv$-type AFQ \label{sec:NMR_100_Qv}}
In the $Q_\varv$-type AFQ state, the effective nuclear Hamiltonian is described as 
\begin{align}
\label{eq:hf_NMR_100_Qv1}
\mathcal{H}_{\rm Co_{A/B}}
=&
\rev{\left( - H_x^{\rm (n)}  + c' M_x^{\rm (c)} \right)  \hat{I}_x
+\left( c_{\rm Q} Q_u^{\rm (c)} \pm cG_{xy}^{\rm (c)} \right)\hat{I}_u
+c' Q_\varv^{\rm (c)} \hat{I}_\varv,} \\
\label{eq:hf_NMR_100_Qv2}
\rev{\tilde{\mathcal{H}}_{\rm Co_{A/B}}
=}&
\rev{\left[ c' \left(Q_u^{\rm (c)} +Q_\varv^{\rm (c)} \pm Q_z^{\rm (c)} \pm   G_{xy}^{\rm (c)} \right) \pm c'_{\rm M} T_y^{\rm (c)} \right]  \hat{I}_x} \notag\\
&\rev{+ c'\left(Q_\varv^{\rm (c)} + M_x^{\rm (c)} \pm Q_z^{\rm (c)} \pm T_y^{\rm (c)} \right) \hat{I}_u} \notag\\
&\rev{+c' \left(Q_u^{\rm (c)} + M_x^{\rm (c)} \pm Q_z^{\rm (c)} \pm G_{xy}^{\rm (c)} \pm T_y^{\rm (c)} \right)\hat{I}_\varv.}
\end{align}

Figure~\ref{fig:figure4}(d) shows the $[100]$-field NMR spectra \rev{for $c=c'=0.02, c'_{\rm M}=0.3$}, where the color scale represents the intensity of the NMR spectra.
The result indicates that sublattice-dependent spectral splitting occurs as well as the results in NQR [Sec.~\ref{sec:NQR_Qv}] and $[001]$-field NMR [Sec.~\ref{sec:NMR_001_Qv}].
\rev{Also in the $[100]$-field NMR, the spectrum is mainly determined by the following dominant contributions: Zeeman term, $c_{\rm Q}Q_u^{\rm (c)}$ term, and primarily induced $G_{xy}^{\rm (c)}$ terms.}
\rev{In other words, a}mong the odd-parity multipoles, $G_{xy}^{\rm (c)}$, $Q_z^{\rm (c)}$, and $T_y^{\rm (c)}$, the important contribution comes from $G_{xy}^{\rm (c)}$, since the magnitudes of $Q_z^{\rm (c)}$ and $T_y^{\rm (c)}$ are much smaller than that of $G_{xy}^{\rm (c)}$, as shown in Fig.~\ref{fig:figure2}(j).
 Meanwhile, the shift of the spectra is dominated by $Q_u^{\rm (c)}$.
 
\subsubsection{Staggered $Q_u$-type AFQ \label{sec:NMR_100_Qu}}
The effective nuclear Hamiltonian in the $Q_u$-type AFQ state is
\begin{align}
\label{eq:hf_NMR_100_Qu1}
\rev{\mathcal{H}_{\rm Co_{A/B}}
=}&
\rev{\left( - H_x^{\rm (n)}  + c' M_x^{\rm (c)} \right)  \hat{I}_x
+ c_{\rm Q} Q_u^{\rm (c)} \hat{I}_u
+\left( c' Q_\varv^{\rm (c)}  \pm cQ_{z}^{\rm (c)}  \right) \hat{I}_\varv,} \\
\label{eq:hf_NMR_100_Qu2}
\rev{\tilde{\mathcal{H}}_{\rm Co_{A/B}}
=}&
\rev{\left[ c' \left(Q_u^{\rm (c)} +Q_\varv^{\rm (c)} \pm Q_z^{\rm (c)} \pm G_{xy}^{\rm (c)} \right) \pm c'_{\rm M} T_y^{\rm (c)} \right]  \hat{I}_x} \notag\\
&\rev{+ c'\left(Q_\varv^{\rm (c)} + M_x^{\rm (c)}\pm G_{xy}^{\rm (c)} \pm Q_z^{\rm (c)} \pm T_y^{\rm (c)} \right) \hat{I}_u} \notag\\
&\rev{+c' \left(Q_u^{\rm (c)} + M_x^{\rm (c)} \pm  G_{xy}^{\rm (c)} \pm T_y^{\rm (c)} \right)\hat{I}_\varv,}
\end{align}
\rev{which is} 
 the same as that in the $Q_\varv$-type AFQ state in Eq\rev{s}.~\eqref{eq:hf_NMR_100_Qv1} \rev{and \eqref{eq:hf_NMR_100_Qv2}}, as the magnetic point group symmetry under the magnetic field is the same as $2'mm'$ with each other.
Thus, in contrast to the results for the NQR [Sec.~\ref{sec:NQR_Qu}] and $[001]$-field NMR [Sec.~\ref{sec:NMR_001_Qu}], the sublattice-dependent splittings occur under the [100] magnetic field as shown \rev{the NMR spectra for $c=c'=0.02, c'_{\rm M}=0.3$} in Fig.~\ref{fig:figure4}(e).

However, the mean-field dependence of the spectra is different from that in the $Q_\varv$-type AFQ state, 
since the magnitude of $Q_z^{\rm (c)}$ is much larger than that of other multipoles.
Especially, the spectral shift reflects the different mean-field dependence of $Q_u^{\rm (c)}$, as already discussed in Sec.~\ref{sec:NQR_Qu}.

\subsubsection{Staggered $M_x$-type AFM \label{sec:NMR_100_Mx}}
The nuclear Hamiltonian in the $M_x$-type AFM state is 
\begin{align}
\label{eq:hf_NMR_100_Mx1}
\rev{\mathcal{H}_{\rm Co_{A/B}}
=}&
\rev{\left( - H_x^{\rm (n)}  + c' M_x^{\rm (c)} \pm c T_y^{\rm (c)}  \right)  \hat{I}_x
+ c_{\rm Q} Q_u^{\rm (c)} \hat{I}_u
+ c' Q_\varv^{\rm (c)}  \hat{I}_\varv,} \\
\label{eq:hf_NMR_100_Mx2}
\rev{\tilde{\mathcal{H}}_{\rm Co_{A/B}}
=}&
\rev{ c' \left(Q_u^{\rm (c)} +Q_\varv^{\rm (c)} \pm Q_z^{\rm (c)} \pm G_{xy}^{\rm (c)} \right)   \hat{I}_x} \notag\\
&\rev{+ c'\left(Q_\varv^{\rm (c)} + M_x^{\rm (c)}\pm G_{xy}^{\rm (c)} \pm Q_z^{\rm (c)} \pm T_y^{\rm (c)} \right) \hat{I}_u} \notag\\
&\rev{+c' \left(Q_u^{\rm (c)} + M_x^{\rm (c)} \pm  G_{xy}^{\rm (c)}  \pm Q_z^{\rm (c)}  \pm T_y^{\rm (c)} \right)\hat{I}_\varv,}
\end{align}
\rev{where the same multipoles appear} 
in the two AFQ states in Eq\rev{s}.~\eqref{eq:hf_NMR_100_Qv1}--\rev{\eqref{eq:hf_NMR_100_Qu2}}, 
since the magnetic point group symmetry under the $[100]$ magnetic field reduces to $2'mm'$ also in this case.
Thus, the sublattice-dependent NMR splittings occur, which is similar to those in the AFQ states. 
However, the dominant odd-parity multipole to induce the spectral splitting is given by $T_y^{\rm (c)}$.
The $[100]$-field NMR spectra \rev{for $c=0.3$ and $c'=0.02$} is shown in Fig.~\ref{fig:figure4}(f).

\section{Spectral splittings under odd-parity multipoles \label{sec:summary_spectra}}
So far, we have focused on the NQR and NMR spectra in the two AFQ and the AFM ordered states under the magnetic fields along the [001] and [100] directions as well as the zero magnetic field. 
In a similar way, possible NQR and NMR splittings in other odd-parity multipole orderings under any field directions can be calculated. 
We show the presence or absence of the sublattice-dependent NQR and NMR splittings for the other candidate odd-parity multipole orders in CeCoSi, which are expected from the low-energy crystal-field level.
\rev{The present analysis is applicable once the phase transition occurs in the magnetic field unless the second excited levels are involved in the phase transition.}
\rev{It is noted that our analysis can be extended to other electronic orderings in different crystal-field levels as discussed in Appendix~\ref{ap:Gamma7} and other multi-orbital systems.}  

\rev{The results in the present $\Gamma_6$-$\Gamma_7$ levels are summarized in Table~\ref{table:summary}.}
\rev{We list} the other candidates\rev{, as discussed in Sec.~\ref{sec:local_multipole}};  
two AFM states, three AFQ states, and two antiferro magnetic octupole (AFO) states.
We also include the results in the $Q_\varv$- and $Q_u$-type AFQ states and the $M_x$-type AFM state discussed in Secs.~\ref{sec:NQR} and \ref{sec:NMR} under the other magnetic-field directions. 
The table exhibits when the sublattice-dependent spectral splittings occur in the presence of odd-parity multipoles. 
\rev{
For example,  in the AFQ phase, the NMR measurement in the $zx$-($yz$-)plane magnetic field is useful to identify the odd-parity multipole order parameter; 
the sublattice-dependent splittings \rev{which always appear when the magnetic field direction is rotated} in the $zx$-($yz$-)plane 
indicate the emergence of $G_{xy}^{\rm (c)}$. 
Meanwhile, in the AFM phase, the sublattice-dependent splittings under the magnetic field along the $x$ direction will 
\rev{indicate the presence of} $T_y^{\rm (c)}$. 
In this way, 
}
\rev{a}s the different spectral splittings are found in the different odd-parity multipole orderings depending on the magnetic field directions, the detailed investigation of the field angle dependence enables us to identify the order parameter in CeCoSi.

\tabcolsep = 1pt
\begin{table}[h!]
\centering
\caption{
The sublattice-dependent NQR and NMR splittings in the AFM, AFQ, and AFO states under the six field directions $[001]$, $[100]$, $[110]$, $_\perp[001]$, $_\perp[010]$, and $_\perp[\bar{1}10]$.
The local multipoles (LMP) at Ce site and cluster odd-parity multipoles (OPMP) are shown in second and third columns, respectively.
The mark $\checkmark$ represents the presence of the sublattice-dependent splittings.
\label{table:summary}}
\vspace{2mm}
\begin{tabular}{ccccccccccc}\hline \hline
 & & & & NQR & \multicolumn{6}{c}{NMR} \\\cline{4-11}
& LMP & OPMP & & - & ${\bm H}_{\parallel[001]}$ & ${\bm H}_{\parallel[100]}$ & ${\bm H}_{\parallel[110]}$ & ${\bm H}_{\perp[001]}$ & ${\bm H}_{\perp[010]}$ & ${\bm H}_{\perp[\bar{1}10]}$ \\\hline
AFM & $M_x$ & $T_y$ & & - &  - & \checkmark & \checkmark & \checkmark & \checkmark & \checkmark \\
& $M_y$ & $T_x$ & &  - &  - & - & \checkmark & \checkmark & - & \checkmark\\
& $M_z$ & $M_u$ &  & - & - & - &  - &  - & \checkmark &  -  \\\hline
AFQ & $Q_u$ & $Q_z$ &  & - & - & \checkmark &  - & \checkmark & \checkmark &  -  \\
& $Q_\varv$ & $G_{xy}$ &  & \checkmark & \checkmark & \checkmark & \checkmark & \checkmark & \checkmark & \checkmark \\
& $Q_{xy}$ & $G_\varv$ &  & - &  - &  - & - & \checkmark &  - &  -  \\
& $Q_{yz}$ & $Q_y$ &  & - &  - & - & - &  - &  - &  \checkmark \\
& $Q_{zx}$ & $Q_x$ &  & - &  - &  - & - &  - & \checkmark & \checkmark \\ \hline
AFO & $M_{xyz}$ & $M_{xy}$ &  & - &  - & - & - &  - &   - & - \\
& $M_z^\beta$ & $M_\varv$ &  & - & \checkmark & - & - &  - & \checkmark & \checkmark  \\
\hline \hline
\end{tabular}
\end{table}

\section{Summary \label{sec:summary}}
We have discussed the effect of the odd-parity multipoles on the NQR and NMR spectra.
First, we have derived the hyperfine field at Co nuclei in consideration of the contribution from the electronic multipole moments at Ce sites. 
We showed that the hyperfine field in the presence of the odd-parity multipole moments cause the sublattice-dependent 
splittings of the NQR and NMR spectra.
Moreover, we obtained the different spectral splittings for the different odd-parity multipoles by considering the NQR spectral splitting as well as $[001]$- and $[100]$-field NMR spectral splittings in the three ordered states, 
the $M_x$-type AFM state with $T_y^{\rm (c)}$ and $Q_\varv$- and $Q_u$-type AFQ states with $G_{xy}^{\rm (c)}$ and $Q_z^{\rm (c)}$\rev{, respectively}.

We emphasize that not only the even-parity multipoles but also the odd-parity multipoles affect the nuclear spin unless the NMR site is located at the inversion center. 
As the key ingredient is the emergence of the cluster odd-parity multipoles, which consist of the spatial distributions of the even-parity multipoles such as magnetic dipole and electric quadrupole, the odd-parity-hosting candidate materials to have the AFM structures, e.g., Ce$_3$TiBi$_5$~\cite{shinozaki2020magnetoelectric} and $A$OsO$_4$($A={\rm K}, {\rm Rb}, {\rm Cs}$)~\cite{yamaura2019crystal}, might be good targeting materials.
\\

\begin{acknowledgments}
We thank M. Manago, K. Kotegawa, H. Tou, H. Tanida, and Y. Ihara for the fruitful discussions on experimental information in CeCoSi.
This research was supported by JSPS KAKENHI Grant Numbers JP18K13488, JP19K03752, and JP19H01834.
M.Y. is supported by a JSPS research fellowship and supported by JSPS KAKENHI (Grant No. JP20J12026).
\end{acknowledgments}

\appendix
\section{Multipole moments under conjugate mean fields  \label{ap:power}}
In this appendix, we discuss the different mean-field dependences of $G^{(\rm c)}_{xy}$ in the $Q_{\varv}$-type AFQ state and $Q^{(\rm c)}_z$ in the $Q_u$-type AFQ state in Sec.~\ref{sec:local_AF}.
The result by the numerical diagonalization indicates that 
$G^{(\rm c)}_{xy}$ roughly increases as a function of $h^{\rm s}_{Q_{\varv}}$, whereas $Q^{(\rm c)}_{z}$ increases as a function of $(h^{\rm s}_{Q_{u}})^3$ in the small $h_X^{\rm s}$ region as shown in Figs.~\ref{fig:figure2}(d) and \ref{fig:figure2}(e).

The difference is understood by the power expansion of the multipole moments, 
\begin{align}
\label{eq:power}
\tilde{X}^{\rm (c)} = X_{\rm A} - X_{\rm B} = a^{(1)}_X (h_{X}^{\rm s}) + a^{(3)}_X (h_{X}^{\rm s})^3 + \cdots,
\end{align}
where $\tilde{X}^{\rm (c)}=G_{xy}^{\rm (c)} (Q_z^{\rm (c)})$ for $X=Q_\varv(Q_u)$.
$a^{(n)}_X$ are the coefficients, which depend on the crystal-field splitting $\Delta$.
It is noted that the even order of $h_{X}^{\rm s}$ does not appear due to the different parity with respect to the inversion symmetry. 

For large $\Delta$, by treating the mean-field term in Eq.~\eqref{eq:local_model} perturbatively, the basis function at Ce$_i$ site in the $Q_\varv$-type AFQ state changes into
\begin{align}
\label{eq:fn_Qv}
\tilde{\phi}_{\Gamma_7\sigma,i} = \frac{1}{N} \left( \phi_{\Gamma_7\sigma,i} \pm \frac{h_{Q_\varv}^{\rm s}}{2\Delta}  \phi_{\Gamma_6\sigma,i} \right),
\end{align}
where the sign $+(-)$ is taken for $i={\rm A}({\rm B})$ and $N$ is the normalization factor.
$\sigma=\uparrow, \downarrow$ is the quasi-spin.
Then, $G_{xy}^{\rm (c)}$ is obtained as
\begin{align}
G_{xy}^{\rm (c)} = \frac{1}{N} \frac{h_{Q_\varv}^{\rm s}}{2\Delta} 
&= \left[1+\left(\frac{h_{Q_\varv}^{\rm s}}{2\Delta }\right)^2 \right]^{-\frac{1}{2}}\frac{h_{Q_\varv}^{\rm s}}{2\Delta}, \notag\\
&\sim \frac{1}{2\Delta}h_{Q_\varv}^{\rm s} - \left( \frac{1}{2\Delta} \right)^3 (h_{Q_\varv}^{\rm s})^3.
\end{align}
This indicates that $a^{(1)}_{Q_\varv}(=\frac{1}{2\Delta}) \gg a^{(3)}_{Q_\varv}[=(\frac{1}{2\Delta})^3]$ is satisfied for $h_{Q_\varv}^{\rm s}/\Delta \ll 1$, 
which results in the linear behavior of $G_{xy}^{\rm (c)}$ in Fig.~\ref{fig:figure2}(d). 
In a similar way, the linear behavior of $T_{y}^{\rm (c)}$ in Fig.~\ref{fig:figure2}(f) is accounted for in the $M_x$-type AFM state.

On the other hand, in the $Q_u$-type AFQ state, $Q_z^{\rm (c)}$ becomes zero for large $\Delta$, which means that
\begin{align}
Q_z^{\rm (c)}&=0 \ (\Delta > h_{Q_u}^{\rm s}),\\
Q_z^{\rm (c)}&=1 \ (\Delta < h_{Q_u}^{\rm s}).
\end{align}
Thus, the onset of $Q_z^{\rm (c)}$ for small $h_{Q_u}^{\rm s}$ in Fig.~\ref{fig:figure2}(e) is owing to the finite temperature effect. 
Numerically, the opposite relation ($a^{(1)}_{Q_u} \ll a^{(3)}_{Q_u}$) to the $Q_u$-type AFQ ordered case with respect to $a^{(1)}_{Q_u}$ and $a^{(3)}_{Q_u}$ is obtained for large $\Delta$; $a^{(1)}_{Q_u} \sim 10^{-2} a^{(3)}_{Q_u}$ for $\Delta=0.5$ and $\beta=10$. 
This implies that $Q^{(\rm c)}_{z}$ increases as a function of $(h^{\rm s}_{Q_{u}})^3$ in the small $h_{Q_u}^{\rm s}$ region in Fig.~\ref{fig:figure2}(e).

\section{Field-swept NMR \label{ap:NMR_H}}
\begin{figure}[htb!]
\centering
\includegraphics[width=88mm]{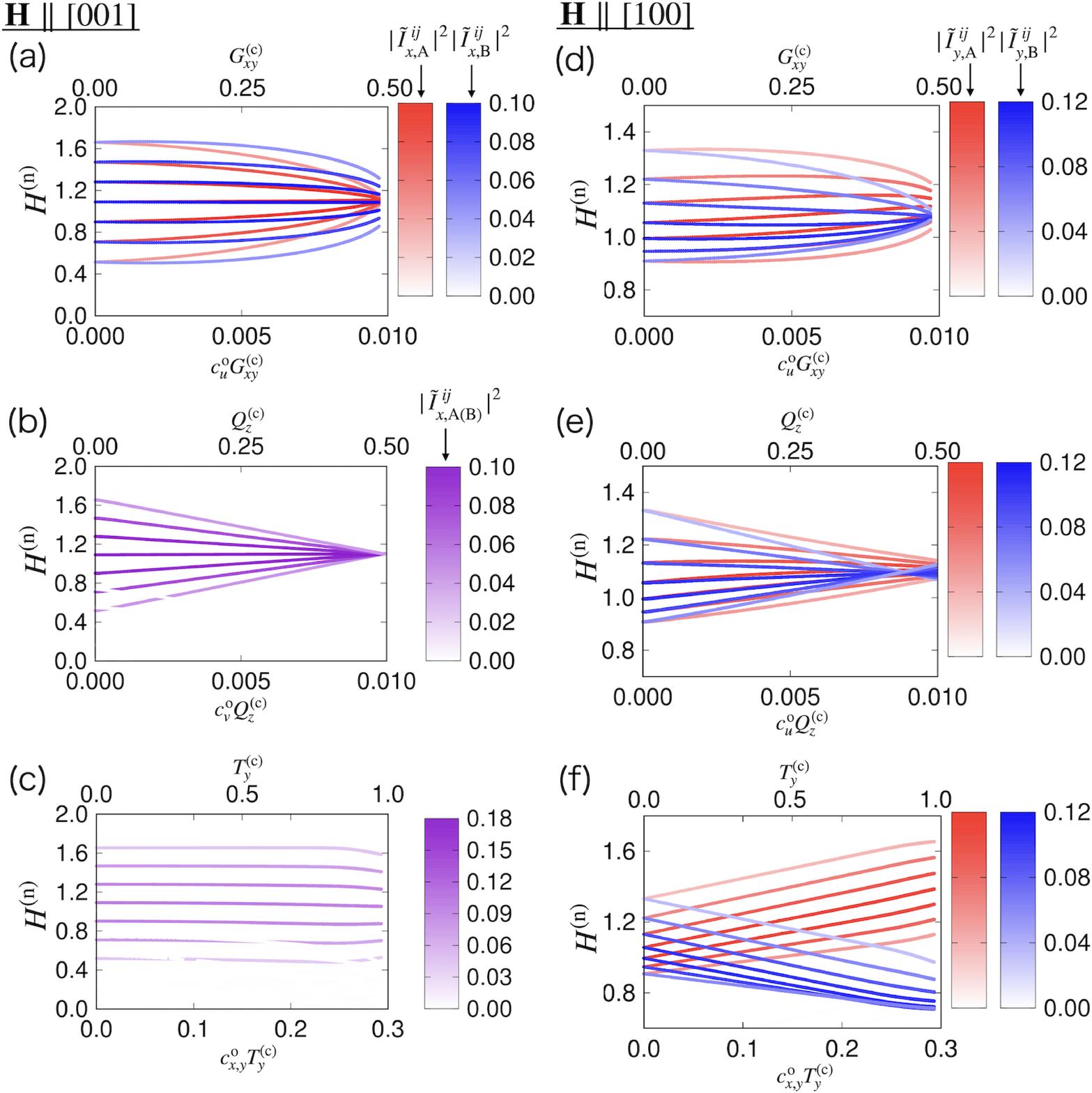}
\caption{The odd-parity multipole dependences of the field-swept NMR spectra at the (a)--(c) $[001]$ magnetic field and (d)--(f) $[100]$ magnetic field.
 The data are for the (a),(d) $Q_{\varv}$-type AFQ, (b), (e) $Q_u$-type AFQ, and (c), (f) $M_x$-type AFM states.
The color scales represent the intensities with (a)--(c) $\left|\tilde{I}_{x, {\rm A(B)}}^{ij}\right|^2$ and (d)--(f) $\left|\tilde{I}_{y, {\rm A(B)}}^{ij}\right|^2$. 
The coupling constants are 
\rev{$c_{u}^{\rm o} = c_{\varv}^{\rm o}=c =0.02$ in the AFQ states and $c_{x,y}^{\rm o}=c=0.3$ in the AFM state.}
\rev{Other coupling constants are set to be $c'=0.02$.}
\label{fig:figureB}}
\end{figure}

We show the field-swept NMR spectra for the resonance frequency $\omega=1.1\gamma$ at the $[001]$ and $[100]$ magnetic fields.
We set $\gamma =1$ and the coupling constant \rev{as well as that in Sec.~\ref{sec:NMR}.} 
Figures~\ref{fig:figureB}(a)--\ref{fig:figureB}(c) show the spectra in the $[001]$ magnetic field, whereas Figs.~\ref{fig:figureB}(d)--\ref{fig:figureB}(f) show those in the $[100]$ magnetic field.
The results show a similar tendency in the cases of the frequency-swept spectra in Fig.~\ref{fig:figure4}. 

\section{$[110]$-field NMR \label{ap:NMR_110}}
We show the effective hyperfine fields and NMR spectra in the case of the $[110]$ magnetic field in the $Q_\varv$- and $Q_u$-type AFQ states and $M_x$-type AFM state.
The hyperfine field Hamiltonian is given by
\rev{
\begin{widetext}
\begin{align}
\label{eq:hf_110_para}
\tilde{\mathcal{H}}_{\rm para}^{[110]} =& 
\left(\tilde{c}_{x,y}^{\rm e,1}Q_u^{\rm (c)} + \tilde{c}_{x,y}^{\rm e,2}Q_{xy}^{\rm (c)} \right) \left(\hat{I}_x + \hat{I}_y\right)   +\left[ \tilde{c}_{u}^{\rm e,1}Q_{xy}^{\rm (c)} + \tilde{c}_{u}^{\rm e,2}\left(M_x^{\rm (c)}+M_y^{\rm (c)}\right)\right]\hat{I}_u 
+\left[\tilde{c}_{xy}^{\rm e,1}Q_u^{\rm (c)} + \tilde{c}_{xy}^{\rm e,2}\left(M_x^{\rm (c)}+M_y^{\rm (c)}\right)\right]\hat{I}_{xy}, \\
\label{eq:hf_110_order_o}
\tilde{\mathcal{H}}_{\rm order}^{{\rm o}[110]} 
=& \tilde{c}_{x,y}^{\rm o,1}G_{xy}^{\rm (c)} \left(\hat{I}_x + \hat{I}_y\right)
+\left[\tilde{c}_{x,y}^{\rm o,2} \left(T_x^{\rm (c)}-T_y^{\rm (c)} \right) +\tilde{c}_{x,y}^{\rm o,3}Q_{z}^{\rm (c)} +\tilde{c}_{x,y}^{\rm o,4}G_\varv^{\rm (c)}\right]\left(\hat{I}_x - \hat{I}_y\right)
+ \tilde{c}_{z}^{\rm o}\left(Q_x^{\rm (c)} - Q_y^{\rm (c)}\right) \hat{I}_z \notag\\
&+ \tilde{c}_{u}^{\rm o}\left(T_x^{\rm (c)}+T_y^{\rm (c)}\right)\hat{I}_u 
+\left[\tilde{c}_{\varv}^{\rm o,1}G_\varv^{\rm (c)} + \tilde{c}_{\varv}^{\rm o,2}\left(T_x^{\rm (c)} - T_y^{\rm (c)}\right)\right]\hat{I}_\varv\notag\\
&+\left[ \tilde{c}_{yz,zx}^{\rm o,1}M_u^{\rm (c)} + \tilde{c}_{yz,zx}^{\rm o,2}M_{xy}^{\rm (c)} \right] \left(\hat{I}_{yz}-\hat{I}_{zx}\right)
+\left[\tilde{c}_{yz,zx}^{\rm o,3}\left(Q_x^{\rm (c)}-Q_y^{\rm (c)}\right)+ \tilde{c}_{yz,zx}^{\rm o,4}M_\varv^{\rm (c)} \right]\left(\hat{I}_{yz}+\hat{I}_{zx}\right)
+\left[\tilde{c}_{xy}^{\rm o,1}G_{xy}^{\rm (c)} + \tilde{c}_{xy}^{\rm o,2}\left(T_x^{\rm (c)} + T_y^{\rm (c)}\right)\right]\hat{I}_{xy}, 
\\
\label{eq:hf_110_order_e}
\tilde{\mathcal{H}}_{\rm order}^{{\rm e}[110]} =&
\left[\tilde{c}_{x,y}^{\rm e,3}\left(M_x^{\rm (c)}-M_y^{\rm (c)}\right)+ \tilde{c}_{x,y}^{\rm e,4}Q_\varv^{\rm (c)}\right] \left(\hat{I}_x - \hat{I}_y\right)+ \left[\tilde{c}_{z}^{\rm e,1}\left(Q_{yz}^{\rm (c)}+Q_{zx}^{\rm (c)}\right) + \tilde{c}_{z}^{\rm e,2}M_{xyz}^{\rm (c)} \right]\hat{I}_z\notag\\
 &+\tilde{c}_{\varv}^{\rm e}\left(M_x^{\rm (c)} -M_y^{\rm (c)}\right)\hat{I}_\varv
+\tilde{c}_{yz,zx}^{\rm e,1}M_z^{\beta{\rm (c)}} \left(\hat{I}_{yz} - \hat{I}_{zx}\right)
+ \left[\tilde{c}_{yz,zx}^{\rm e,2}\left(Q_{yz}^{\rm (c)} + Q_{zx}^{\rm (c)}\right) +\tilde{c}_{yz,zx}^{\rm e,3}M_z^{\rm (c)} + \tilde{c}_{yz,zx}^{\rm e,4}M_{xyz}^{\rm (c)} \right]\left(\hat{I}_{yz}+\hat{I}_{zx}\right).
\end{align}
\end{widetext}
}

We set the coupling constants as 
\rev{$c_{u}^{\rm e} = c_{\rm Q}=0.13$,}
\rev{$c_{x,y}^{\rm o}=0.3$, }
and the others are set to be \rev{$0.02$} for simplicity.

\begin{figure}[htb!]
\centering
\includegraphics[width=88mm]{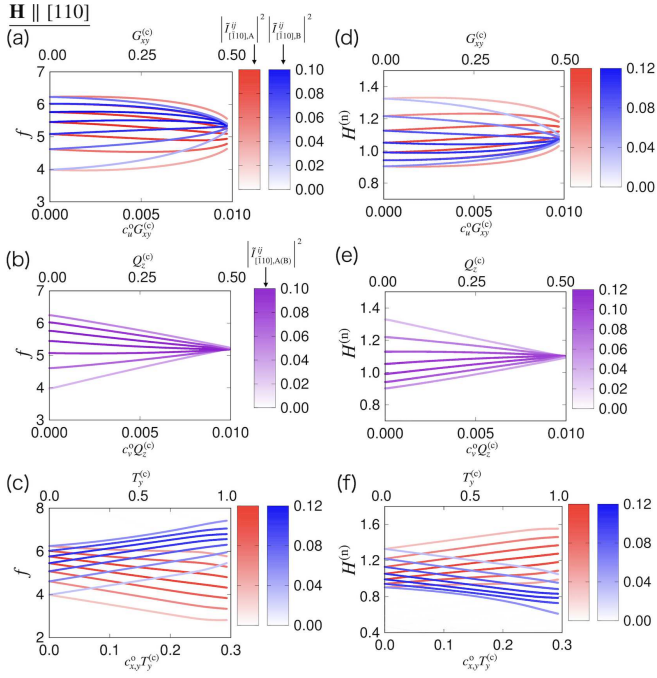}
\caption{
The odd-parity multipole dependences of the (a)--(c) frequency-swept NMR spectra and (d)--(f) field-swept NMR spectra under the $[110]$ magnetic field. 
The data are for the (a),(d) $Q_{\varv}$-type AFQ, (b), (e) $Q_u$-type AFQ, and (c), (f) $M_x$-type AFM states.
The color scales represent the intensities with $\left|\tilde{I}_{[\bar{1}10], {\rm A(B)}}^{ij}\right|^2$. 
The coupling constants are \rev{$c_u^{\rm o} = c_{\varv}^{\rm o} = 0.02$ in the AFQ states and $c_{x,y}^{\rm o}=0.3$ in the AFM state.}
\rev{Other coupling constants are set to be $c'=0.02$.}
\label{fig:figureC}}
\end{figure}

Figures~\ref{fig:figureC}(a)--\ref{fig:figureC}(c) show the frequency-swept NMR spectra for the magnetic field $|{\bm H}^{\rm (n)}| =1$, 
whereas Figs.~\ref{fig:figureC}(d)--\ref{fig:figureC}(f) are the field-swept NMR spectra for the resonance frequency $\omega=1.1\gamma$, where $\gamma$ is set to be $1$.
The intensity of the spectra is calculated by $\left|\tilde{I}_{[\bar{1}10], {\rm A}({\rm B})}^{ij}\right|^2$ for $I_{[\bar{1}10]}=(I_x-I_y)/2$.

In the $Q_\varv$-type AFQ [Figs.~\ref{fig:figureC}(a) and \ref{fig:figureC}(d)] and $M_x$-type AFM states [Figs.~\ref{fig:figureC}(d) and \ref{fig:figureC}(f)], the splittings in the [110] field show a similar tendency to those in the [100] field in Secs.~\ref{sec:NMR_100_Qv} and \ref{sec:NMR_100_Mx}.
Their splittings are dominantly characterized by $G_{xy}^{\rm (c)}$ and $T_{y}^{\rm (c)}$, respectively.
On the other hand, in the $Q_u$-type AFQ state in Figs.~\ref{fig:figureC}(b) and \ref{fig:figureC}(e), there are no spectral splittings in contrast to the result under the $[100]$ field in Secs.~\ref{sec:NMR_100_Qu}.
The reason why no splittings occur under the [110] field is attributed to the difference of the site symmetry at Co site.
As the present site symmetry is $2'22'$, which is different from $2'mm'$ in the $[100]$ direction, there is no 
coupling between odd-parity $Q_z^{\rm (c)}$ and any of $I_x+I_y$, $I_u$, and $I_{xy}$ in Eq.~\eqref{eq:hf_110_order_o}.

\section{Spectral splittings for another low-energy crystal-field levels \label{ap:Gamma7}}

The different low-energy crystal-field levels activate different types of odd-parity multipole orderings.
In this appendix, we show the expected sublattice-dependent splittings in NQR and NMR spectra by supposing the low-energy crystal-field level consisting of the two $\Gamma_7$ doublets~\cite{nikitin2020gradual}.
In this case, other two multipole orderings become possible: $Q_{4z}^\alpha$-type antiferroic hexadecapole ordering (AFH)  with the odd-parity electric toroidal quadrupole $G_u$ and $M_{5u}$-type antiferroic triacontadipole ordering (AFT) with the magnetic toroidal dipole $T_z$, where the functional forms of $Q_{4z}^\alpha$ and $M_{5u}$ are shown in Ref.~\onlinecite{hayami2018classification}.

By performing a similar procedure in Secs.~\ref{sec:hf},~\ref{sec:NQR}, and~\ref{sec:NMR}, the presence or absence of the sublattice-dependent spectral splittings in NQR and NMR is obtained.
The results are summarized in Table~\ref{table:summary_77}.
The common multipoles appearing in both the two $\Gamma_7$ doublets and $\Gamma_6$-$\Gamma_7$ doublets, $T_x$, $T_y$, $M_u$, $Q_z$, $Q_x$, and $Q_y$, give the same result in Table~\ref{table:summary}.
Note that electric toroidal quadrupole $G_\varv, G_{xy}$ and magnetic quadrupole $M_\varv, M_{xy}$ are not activated within the low-energy crystal-field levels unless the first-excited state is $\Gamma_6$ doublet.

\tabcolsep = 1pt
\begin{table}[htb!]
\centering
\caption{
The sublattice-dependent NQR and NMR splittings in the AFM, AFQ, AFH, and AFT states under the six field directions $[001]$, $[100]$, $[110]$, $_\perp[001]$, $_\perp[010]$, and $_\perp[\bar{1}10]$, when the crystal-field first-excited state is $\Gamma_7$ doublet.
The local multipoles (LMP) at Ce site and cluster odd-parity multipoles (OPMP) are shown in second and third columns, respectively.
The mark $\checkmark$ represents the presence of the sublattice-dependent splittings.
\label{table:summary_77}}
\vspace{2mm}
\begin{tabular}{ccccccccccc}\hline \hline
 & & & & NQR & \multicolumn{6}{c}{NMR} \\\cline{4-11}
& LMP & OPMP & & - & ${\bm H}_{\parallel[001]}$ & ${\bm H}_{\parallel[100]}$ & ${\bm H}_{\parallel[110]}$ & ${\bm H}_{\perp[001]}$ & ${\bm H}_{\perp[010]}$ & ${\bm H}_{\perp[\bar{1}10]}$ \\\hline
AFM & $M_x$ & $T_y$ & & - &  - & \checkmark & \checkmark & \checkmark & \checkmark & \checkmark \\
& $M_y$ & $T_x$ & &  - &  - & - & \checkmark & \checkmark & - & \checkmark\\
& $M_z$ & $M_u$ &  & - & - & - &  - &  - & \checkmark &  -  \\\hline
AFQ & $Q_u$ & $Q_z$ &  & - & - & \checkmark &  - & \checkmark & \checkmark &  -  \\
& $Q_{yz}$ & $Q_y$ &  & - &  - & - & - &  - &  - &  \checkmark \\
& $Q_{zx}$ & $Q_x$ &  & - &  - &  - & - &  - & \checkmark & \checkmark \\ \hline
AFH & $Q_{4z}^\alpha$ & $G_u$ &  & - &  - &  - & \checkmark & \checkmark &  - &  \checkmark  \\
\hline
AFT & $M_{5u}$ & $T_z$ &  & - &  - & - & - &  - &   - & \checkmark \\
\hline \hline
\end{tabular}
\end{table}

\bibliographystyle{apsrev4-2}
\bibliography{ref}

\begin{thebibliography}{78}
\bibitem [{M_c()}]{M_comment}%
  \BibitemOpen
  \href@noop {} {}\bibinfo {note} {There are two other magnetic multipole
  degrees of freedom consisting of $x$ or $y$ component of two intraorbital and
  an interorbital magnetic dipoles, which are orthogonal with each other. For
  $z$ component, there is another magnetic multipole degree of freedom from two
  intraorbital magnetic dipoles.}\BibitemShut {Stop}%
\bibitem [{Q_c()}]{Q_comment}%
  \BibitemOpen
  \href@noop {} {}\bibinfo {note} {As well as magnetic dipole case, there is
  another electric multipole degree of freedom from $\hat{Q}_0^{\rm (c)}$ and
  $\hat{Q}_u^{\rm (c)}$ in the same irreducible representation with
  $\hat{Q}_z^{\rm (c)}$.}\BibitemShut {Stop}%
\bibitem [{Del()}]{Delta_comment}%
  \BibitemOpen
  \href@noop {} {}\bibinfo {note} {We also avoid the situation where some
  multipole moments, such as $Q_{yz}, Q_{zx}, Q_{xy}, M_{xyz}, T_x, T_y, M_u$,
  and $M_{xy}$, vanish by taking specific values,
  $\delta^{\Gamma_6}=\delta'=1$, in the $Q_u$-type AFQ state for the magnetic
  field in the plane normal to the $[\bar{1}10]$ direction.}\BibitemShut
  {Stop}%
\bibitem [{CCF()}]{CCF_comment}%
  \BibitemOpen
  \href@noop {} {}\bibinfo {note} {The following result does not change for
  nonzero $C_\nu^{\rm CF}$ in the present model.}\BibitemShut {Stop}%
\end{thebibliography}

\end{document}